\documentclass[]{aa}
\usepackage{graphicx}
\usepackage{txfonts}
\usepackage{natbib}
\usepackage{graphicx}
\usepackage{hyperref}
\usepackage{booktabs}
\usepackage{multicol}
\usepackage{tabularx}
\usepackage[table]{xcolor}
\usepackage{amsmath}
\usepackage{amssymb}
\usepackage{mathtools}
\usepackage{caption}
\usepackage{subcaption}
\usepackage{txfonts}
\usepackage{makecell}
\usepackage{indentfirst}
\usepackage{longtable}
\usepackage{algorithmic}
\usepackage[lined,ruled,linesnumbered]{algorithm2e}
\usepackage[table, dvipsnames]{xcolor}
\usepackage{nicematrix}
\NiceMatrixOptions{cell-space-top-limit=2pt,cell-space-bottom-limit=1.5pt}
\usepackage{comment}
\usepackage[normalem]{ulem}
\usepackage{multirow,booktabs}
\usepackage[flushleft]{threeparttable} 
\usepackage{caption}
\usepackage{adjustbox}
\usepackage{threeparttable}

\newcommand\gaia{\textit{Gaia}\xspace}

\newcommand\gdrthree{\gaia DR3\xspace}

\newcommand\gdrfour{\gaia DR4\xspace}

\newcommand\gfpr{\gaia FPR\xspace}

\usepackage{supertabular}

\definecolor{myviolet}{RGB}{245,200,245} % very light violet

\definecolor{myblue}{RGB}{180,200,255} % very light violet

\newcommand{\swBH}{713\xspace}
\newcommand{\mwBHfisher}{588\xspace}
\newcommand{\mwBHminp}{605\xspace}
\newcommand{\mwBH}{735\xspace}

\newcommand{\bt}{{\bf{t}}}
\newcommand{\bn}{{\bf{n}}}
\newcommand{\by}{{\bf{y}}}

\newcommand{\bM}{{\bf{M}}}

\newcommand{\bgamma}{{\boldsymbol{\gamma}}}
\newcommand{\bsigma}{{\boldsymbol{\sigma}}}

\newcommand{\bbet}{{\boldsymbol{\beta}}}

\newcommand{\bcc}{{\bf{c}}}
\newcommand{\bss}{{\bf{s}}}

\definecolor{mulberry}{rgb}{0.77, 0.29, 0.55}

\makeatletter
\renewcommand*\aa@pageof{, page \thepage{} of \pageref*{LastPage}}
\makeatother

\setcitestyle{citesep={,}}

\newcommand{\orcit}[1]{\protect\href{https://orcid.org/#1}{\protect\includegraphics[width=8pt]{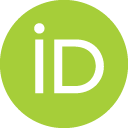}}}

\makeatletter
\renewcommand*\maketitle{%
  \thispagestyle{firstpage}
\begingroup
    \if@wideboxfn
    \setlength\bibindent{1.4\parindent}
    \else
    \setlength\bibindent{\parindent}
    \fi
    \renewcommand*\thefootnote{\@fnsymbol\c@footnote}%
    \renewcommand\@makefntext[1]{%
    \ifaa@longfn\hsize\textwidth\fi
    \noindent
    \hb@xt@\bibindent{\hss\@makefnmark\enspace}##1}
  \ifaa@twocolumn
  \begingroup
    \begin{aa@strip}
          \aa@maketitle
    \end{aa@strip}
    \@thanks            
  \endgroup
  \else
    \begingroup
      \let\thanks\footnote
      \aa@maketitle
    \endgroup
  \fi
\endgroup
  \setcounter{footnote}{0}%
}
\makeatother

\begin{document}
\title{Follow the wobble: Statistical methods to detect astrometric binary asteroids in \gfpr}
\titlerunning{Follow the wobble in \gfpr}

\author{
L. Liberato     \orcit{0000-0003-3433-6269}\inst{\ref{inst:0001}}
\and
P. Tanga        \orcit{0000-0002-2718-997X}\inst{\ref{inst:0001}}
\and
D. Mary         \orcit{0000-0002-9047-5768}\inst{\ref{inst:0001}}
\and
R. Lallemand    \orcit{0009-0009-2967-6948}\inst{\ref{inst:0006}}
\and
Z. Liu            \orcit{0009-0004-6815-5794}\inst{\ref{inst:0006}}
\and
B. Carry        \orcit{0000-0001-5242-3089}\inst{\ref{inst:0001}}
\and
J. Desmars      \orcit{0000-0002-2193-8204}\inst{\ref{inst:0007},\ref{inst:0006}}
\and
D. Hestroffer        \orcit{0000-0003-0472-9459}\inst{\ref{inst:0006}}
\and 
K. Minker      \orcit{0009-0002-6435-9453}\inst{\ref{inst:0009}}
%\and D. Oszkiewicz   \orcit{0000-0002-5356-6433}\inst{\ref{inst:0005}}
%\and B. Sicardy      \orcit{0000-0003-1995-0842}\inst{\ref{inst:0004}}
%\and F. Spoto        \orcit{0000-0001-7319-5847}\inst{\ref{inst:0003}}
\and
A. Siakas \orcit{0009-0004-4243-0372}\inst{\ref{inst:0010}}
}

%P. Bartczak     \orcit{0000-0002-3466-3190}\inst{\ref{inst:0005},\ref{inst:0008}}

\institute{ 
Université Côte d’Azur, Observatoire de la Côte d’Azur, CNRS, Laboratoire Lagrange, Bd de l’Observatoire, CS 34229, 06304 Nice Cedex 4, France\relax \label{inst:0001}
%\and Harvard-Smithsonian Center for Astrophysics, 60 Garden St., MS 15, Cambridge, MA 02138, USA\relax \label{inst:0003}
%\and LESIA, Paris Observatory, PSL University, CNRS, Sorbonne University, Univ. Paris Diderot, Sorbonne Paris Cité, 5 place Jules Janssen, 92195 Meudon, France\relax\label{inst:0004}
%\and Astronomical Observatory Institute, Faculty of Physics, Adam Mickiewicz University, Słoneczna 36, 60-286 Poznań, Poland\relax\label{inst:0005}
\and Laboratoire Temps Espace (LTE), Observatoire de Paris, Université PSL, Sorbonne Université, 77 avenue Denfert Rochereau 75014 Paris, France\relax \label{inst:0006}
\and Polytechnic Institute of Advanced Sciences-IPSA, 63 Boulevard de
Brandebourg, 94200 Ivry-sur-Seine, France\relax \label{inst:0007}
%\and Instituto Universitario de Física Aplicada a las Ciencias y las Tecnologías (IUFACyT). Universidad de Alicante, Ctra. San Vicente del Raspeig s/n. 03690 San Vicente del Raspeig, Alicante, Spain\relax\label{inst:0008}
\and Lowell Observatory, 1400 Mars Hill Rd. Flagstaff, Arizona 86001, USA\relax \label{inst:0009}
\and Department of Physics, Aristotle University of Thessaloniki,  
University Campus, Thessaloniki, 54124, Greece\relax \label{inst:0010}
}

\date{Accepted}

\abstract
{In a previous article, thanks to \gdrthree astrometric accuracy, we obtained the first-ever list of astrometric binary asteroid candidates. Some of these candidates have now been  confirmed. In that previous work, however, the details of the statistical methods were not provided.}
{Our first aim is to provide methodological  details and performance evaluation of the  approach used for detecting binaries. Our second aim is to establish an updated list of binary asteroid candidates from \gfpr astrometric residual exploration, accounting for the statistical properties of the \gfpr data.}
{We account for the astrometric uncertainties from \gfpr and we refine the statistical model of the data, which we use in MC simulation to evaluate the strength of the individual detections; we set up a  trend detection method in the residuals and apply a dedicated period search algorithm; we update the statistical selection process to build the list of candidates; we set up a method for detecting objects in multiple windows of consecutive observation; we refine the method for confidence interval estimation of these parameters and we better constrain the physical parameter selection.}
{We detect 343 binary asteroid candidates corresponding to 410 windows of consecutive observations in the  astrometric data. We show that in noise-only control simulations, the typical number of detections is 88\% lower than in the \gfpr data. We also detect 9 known binaries, 25 candidates overlapping with the Pan-STARRS survey, and 99 candidates overlapping with our previous binary search in \gdrthree. Finally, we report the detection of 45 objects with trends in residuals suggestive of wide binary systems.}
{Our results and analyses demonstrate that although detecting binary asteroids is a difficult problem due to their low signal level, the proposed method is likely to provide a reliable list of detections, including systems poorly accessible to conventional techniques. This set of targets is valuable for future confirmation with stellar occultations, light curves, and forthcoming LSST data.}

\maketitle

%*******************************************************************************************************************************************************************************************************************************************************

\section{Introduction}
\label{sec:intro}
Asteroids are a gold mine of information. They can tell us, for instance, the material composition and distribution in the protoplanetary disk  \citep{carry2012density,demeo2015compositional}, the processes of the planetary formation \citep{kleine2002rapid,izidoro2015terrestrial}, and the evolution of the Solar System until its current state \citep{demeo2014solar,morbidelli2015dynamical}. Binary asteroids are the easiest way to gather this information since they are small-scale laboratories of planetary formation, and some of them may carry the footprints from the primordial Solar System. However, discovering binary asteroids is not easy, which can be seen by the current small number of known binary systems, when it is expected to represent about 20\% of the asteroid population \citep{pravec2007binary,margot2015}. 

In a previous article (\cite{liberato2024}, hereafter L24), we presented the results of a new method to discover binary asteroids in the Solar System by detecting their astrometric wobble, i.e. the periodic variations in the astrometry due to the gravitational perturbation from a companion. We explored the astrometric data available in \gaia Data Release 3 (DR3) for more than 150,000 asteroids during 34 months of operation. As a result, we obtained a list with more than 350 binary asteroid candidates. With the publication of our first astrometric binary exploration, several objects in our list had their binary nature confirmed, such as (3220) Murayama \citep{Benishek2025,Sato2013}, (720) Bohlinia \citep{gorshanov2025}, (1879) Broederstroom \citep{Benishek2024a}, (1967) Menzel \citep{monteiro2024}%,monteiro2026}
, which shows that our method was in fact useful to detect astrometric signals from binary asteroids. Additionally, for other objects, results from stellar occultations have shown ambiguous evidence of their binarity, indicating the possibility of a contact binary \citep{lallemand2025sf2a%,lallemand2026
}. In the present paper, we aim to provide methodological details for searching binary asteroids and apply it to the \gaia Focused Product Release (FPR) \citep{gaiafpr} astrometric data, which contains the same $\sim$152,000 objects as in \gdrthree but with the observations spanning over 66 months. The approach used in L24 can be summarised as follows:
\begin{itemize}
    \item Our data sample consisted of transit--averaged residuals from the orbital fitting of the astrometric data from \gdrthree projected in the along scan (AL) direction of \gaia;
    \item The uncertainties affecting these residuals were taken as the standard deviation of the residuals per transit, also projected in AL direction;
    \item To avoid changes in the geometry of observations that could drastically affect the phase and amplitude of the astrometric wobble detection, we performed the period search in windows of observations (WO) that contained at least 10 transits in a maximum of 10-day span\footnote{These parameters were adopted in L24 as a compromise between the observation arc and the number of exploitable targets.};
    \item We  used Generalised Lomb-Scargle Periodogram \citep[GLSP, ][]{vanderplas2018,Lomb1976,Scargle1982} to run a period search in all of the WOs, considering the largest peak of the periodogram as the potential signal;
    \item We estimated a ``significance'' for the peak value by computing its p-value, which is an indicator of how likely it would be to find a similar peak value in a noise-only situation;
    \item We also estimated confidence intervals for the signal period and amplitude, and computed a so-called  ``quality factor Q'' also aimed at measuring the detection strength;
    \item  We then selected the candidates having p-values smaller than 5\% and Q factors larger than 50\%;
    \item The last step was to evaluate the estimated parameters' coherence with the binary asteroid model adopted through the estimation of the minimum density of the objects and minimum separation between the components of the binary candidates. 

\end{itemize}

When updating this method for \gfpr data, we find that the  approach above could  be improved and optimised in several ways, namely : the use of \gfpr data at the CCD level, in contrast to that of \gdrthree, offer the opportunity to better account for the errors bars when computing data at the transit level; the reliability of these errors could be evaluated, and propagated to estimate the uncertainties  at the transit level; the method for confidence intervals (CI) estimation was improved, with now reduced false coverage; finally, the criterion for selecting the most interesting candidates (lowest p-values) was changed from a simple threshold to a classical algorithm aimed at  controlling the false discovery rate.

In Sec.~\ref{sec:errors}, we discuss the new error model adopted; We present the challenges on the CI estimation and the algorithm implementation for this work in Sec.~\ref{sec:ci_est}; In Sec.~\ref{sec:trendy}, we show an intriguing trendy behaviour observed in the residuals and how we explore it in the detection of binaries; We then present the new approach used on the statistical selection of objects in Sec.~\ref{sec:stat_sel}; We discuss the updates in the physical validation of the outcomes from the previous step in Sec.~\ref{sec:physical}; We apply our revised approach to the astrometric data in \gfpr and present our results and discussions in Sec.~\ref{sec:results}. Finally, we draw our conclusions in Sec.~\ref{sec:concl}.

%*******************************************************************************************************************************************************************************************************************************************************
\section{Noise model}
\label{sec:errors}
In this section, our goal is to present a statistical model of the data that accurately propagates the error properties from the observation (CCD level) to the transit level. This statistical model is critical to ensure that the data exploitation is reliable, as it directly impacts the determination of the p-values used for the detection of the candidates, and the design of reliable CIs for the amplitude and period of the detected binaries. We first recall in Sec.~\ref{subsec:prop_gaia} the properties of the astrometry data and show that the random error provided by \gfpr for astrometry can be used to estimate accurately the standard deviation (hereafter std) of the error at CCD level on the residuals in the AL direction.
We then present in Sec.~\ref{subsec:model} the statistical data model per observation and at the transit level, along with the noise parameters and how those are computed. Throughout the paper, bolded letters denote column vectors.

\subsection{Properties of the \gaia astrometric data}
\label{subsec:prop_gaia}

All main features of \gaia asteroid astrometry have been thoroughly described in previous publications \citep{gaiafpr, tanga2023, gaiacollaborationDR2_2018}. Here, we just briefly recall the main properties that directly affect their exploitation in our case:
\begin{itemize}
    \item \gaia observes targets' positions on the focal plane, which correspond to independent measurements on different CCDs;
    \item Each CCD position measurement is referred to as an ``observation''. The observations are grouped by ``transits'', which happen over a short time span ($\approx 40~s$). There can be a maximum of N = 9 positions per transit, but for minor bodies, this is often not the case; 
    \item Transits are spaced in time irregularly, with long periods without any observation for a given target. Sequences of consecutive transits (minimum interval, 106 minutes) are common, but their frequency decreases with their length.
    \item The maximum accuracy provided by \gaia is in the along scan direction (AL), so that measurements can be considered as essentially one-dimensional.
    \item The AL direction changes gradually its orientation over time (with the precession of the satellite spin axis), so that any direction on the sky is scanned, over time, with different orientations of AL. The same applies to moving sources.
    \item The error model of each observation must take into account the direction of AL and all dependencies of the derived positions on errors coming from the calibration of the focal plane, the reconstructed attitude of the satellite, etc. We distinguish two main components in the final error: a systematic component, assumed to be constant along a transit and common to all observations within the transit; a random component, different for individual observations \citep{lindegren2021,tanga2023}.  
\end{itemize}

As will be discussed in the next section, the systematic component (called $\mu$) cannot be disentangled from the wobble signature because both act as a common shift on the observations over the same transit. However,  wobbles can indeed be detected when combining several transits over a window because the signal changes over time\footnote{In fact, the systematic component may also change from one transit to another, so wobbles can be detected if the systematic components do not vary much, and/or if it is small compared to the wobble amplitude.}. As for the random component, estimating its variance accurately is not straightforward.  In L24, we used the empirical variance directly from the post-fit residuals in the AL direction obtained from the $N$ observations of the transit. While this is a classical estimate, it is also significantly noisy because $N$ is small ($N\leq 9$) and the variance of the estimator of the variance estimate decreases as $\frac{1}{N}$. 

\begin{figure}[htpb]
    \centering
    \includegraphics[width=\linewidth]{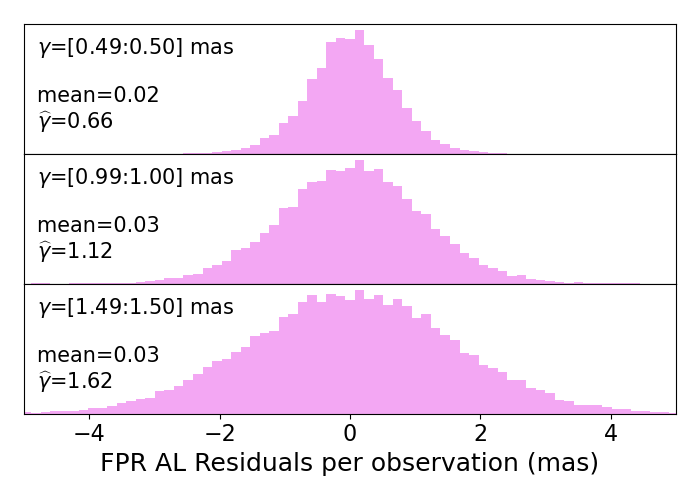}
    \caption{Distribution of AL projection of residuals from \gfpr per observation, i.e. not averaged by transit. For each plot, we chose an interval of random errors per observation projected in AL ($\gamma$) from \gfpr. For each observation in the \gfpr catalogue having $\gamma$ within this interval, we retrieve the corresponding residual, and the distribution of such residuals is plotted in pink, with their empirical mean and the std $\widehat{\gamma}$ indicated. The distributions in the top, middle and bottom plots contain, respectively, 22k, 45k and 56k observations each.}
    \label{fig:fpr_error}
\end{figure}

However, an estimate of the std of the random component in RA and DEC (called $\widehat{\gamma}$ below) is provided in \gfpr data, which can be projected in the AL direction. 
To check whether this projected std is a good proxy to the actual std of the observation data, we selected AL residuals per observation having a value of $\widehat{\gamma}$ in a given narrow range (one per panel in Fig.~\ref{fig:fpr_error}), plotted the resulting distributions and computed the empirical std ($\widehat{\gamma}$ in the legends).

We can see that the empirical dispersion of the AL residuals (per observation) from \gfpr is larger, though marginally, than the ``theoretical'' error $\widehat{\gamma}$ in AL direction as provided by the \gfpr error model. 
There are several reasons for the slight underestimation observed here. First, the residuals from the orbital fit include the systematic components: this spreads the empirical distribution, as it acts as a random mean added to the residuals. However, this is not the only effect\footnote{The std of the systematic component over all post-fit residuals (say, $\widetilde{\mu}$) per transit $\tilde{\mu}$ was estimated as $0.27$ mas, see L24. For the first panel, for instance, if this effect were the only cause, we should obtain $\widehat{\gamma}=~\sqrt{\gamma^2+0.27^2}\approx0.56$ mas, which is smaller than the observed value $\widehat{\gamma}=~0.66$ mas. The mismatch is similar in the other dispersion ranges.}.
For instance, recent systematic exploitation has shown that differences between photocentre and barycentre should be taken into account for the most accurate orbital fit, especially for bright asteroids \citep{fuentes2024}, so such differences may create additional scatter. In conclusion,  although there exists some slight disagreement between the residual's dispersion and the theoretical uncertainty provided by the \gaia error model, our investigations (Fig.~\ref{fig:fpr_error}) indicate that the value provided by \gfpr fairly reflects the actual dispersion, and we will use it in the subsequent processing stages.

\subsection{Data model and parameter estimation}
\label{subsec:model}
We first describe the data model per observation, i.e., at the CCD level, and then turn to the data at the transit level, which are a combination of $N$ observations' measurements. Let ${\bf{t}}:=~[t_1,\cdots, t_N]^\top$ be the vector with epochs of the observations over one transit, $r$ be the vector of residuals, and let $r_i:=~r(t_i)$ be the residual for a given observation obtained from the difference between the astrometric positions of the asteroid and the orbital fit to these measurements, projected in the AL direction.

These residuals are stored in vector ${\bf{r}}:=~[r_1,\cdots, r_N]^\top$. 
As discussed in Sec.~ \ref{subsec:prop_gaia}, the error on the astrometric measurement has two components : an unknown systematic offset, noted $\mu$ below,  and a random  perturbation, modelled as a Gaussian noise  ($n$), with variance   $\gamma^2$ provided by \gfpr ( Sec.~\ref{subsec:prop_gaia}). The random perturbation on one observation is noted $n_i:=~n(t_i)$  with $n_i\sim {\cal{N}}(0,\gamma^2_i)$ where $\gamma^2_i:=~\gamma^2(t_i)$.
This leads to the following data model for the residuals per observation:

\begin{equation}\label{eq:StatModel2}
    \begin{cases}
        \mathcal{H}_0: r_i=  \mu + n_i, \\
        \mathcal{H}_1: r_i =  \mu + A \sin(2\pi f t_i + \varphi) + n_i,
    \end{cases}
\end{equation}

where ${\mathcal{H}_0}$ denotes the hypothesis where no wobble is present in the data, and ${\mathcal{H}_1}$ denotes the alternative hypothesis (wobble present).
Under $\mathcal{H}_1$,  the terms $A$, $f$ and $\varphi$ denote respectively the unknown amplitude, frequency and phase of the sinusoid that models the binary wobble in the AL direction. Note that the wobble   period is much longer than the transit duration ($\approx 40~s$), so that this term can be considered constant over a transit.

\begin{figure}[htpb]
    \centering
    \includegraphics[width=\linewidth]{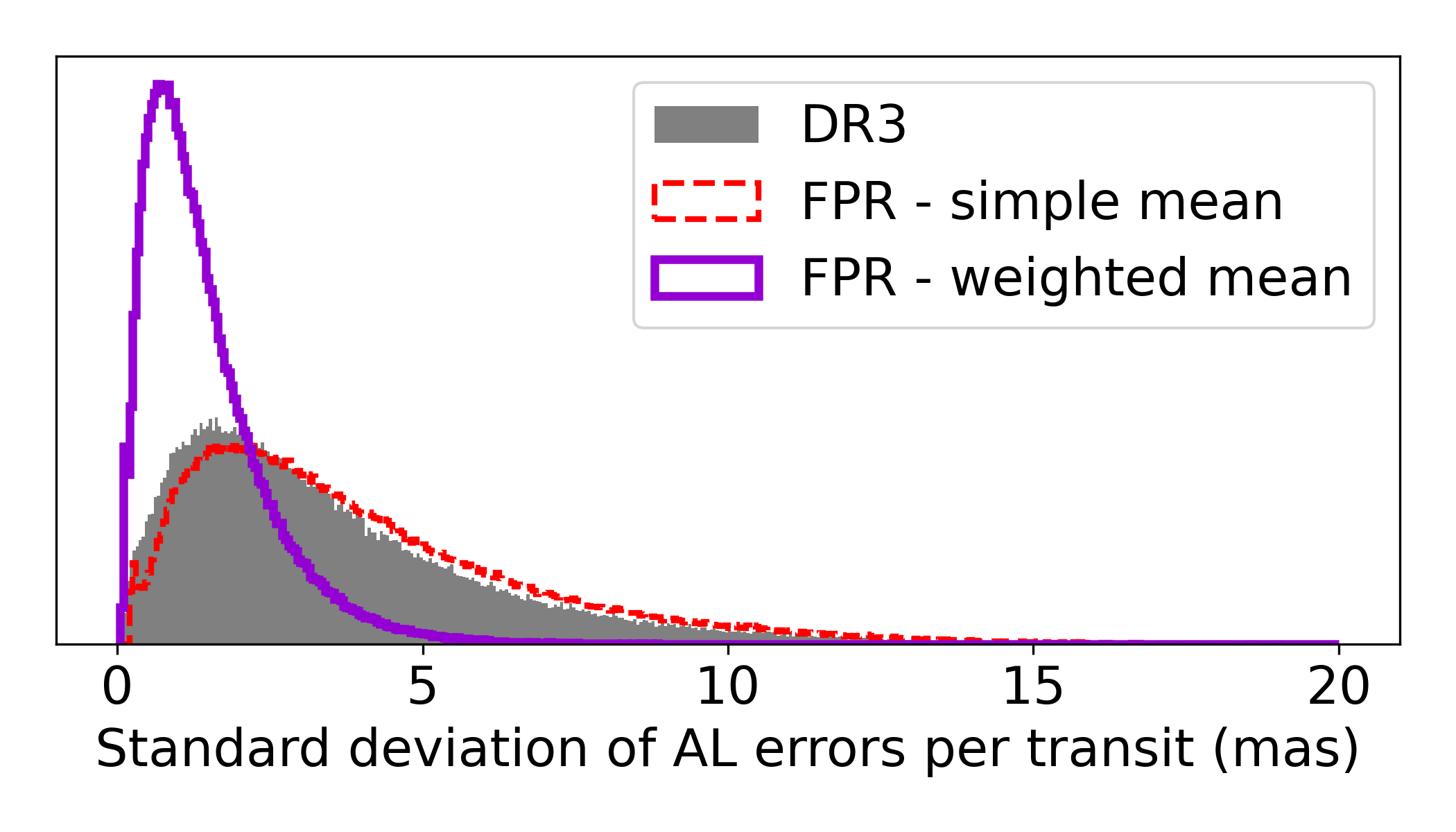}
    \caption{Comparison between the distribution of 50,000 values of standard deviation per transit projected in AL direction, obtained when computing the transit data in different ways.}
   
    \label{fig:err_dr3_fpr}
\end{figure}

From model \eqref{eq:StatModel2}, our first goal is to extract the information about the constant term from $N$ measurements $\{r_i\}_{i=1,\cdots, N}$, to produce an estimate of the residual at the transit level (noted $y$ below), which contains the wobble signature under $\mathcal{H}_1$. Standard Maximum Likelihood Estimation of the constant term in model \eqref{eq:StatModel2} leads to

\begin{equation}
    y=\displaystyle{\sum_{i=1}^N \frac{ r_{i}/\gamma_i^2}{1/\gamma_i^2}}
\end{equation}
It is easy to see that this estimate is Gaussian, unbiased (with mean the constant term) and has std $\overline{\gamma}$ given by:
\begin{equation}\label{gammabar}
    \overline{\gamma}=\displaystyle \sqrt{\frac{1}{\sum_{i=1}^N{1/\gamma_i^2}}}.
\end{equation}

Let us now denote by $k$ the index of the transit to which the $N$ observations per transit belong, and by $t_k$ the corresponding transit epoch\footnote{$t_k$ appears only in the sinusoidal term; since it is essentially constant over the transit with respect to the duration of a wobble, it can be taken as the mean of the $N$ epochs, or as any of the $t_i$ since this has negligible impact on the value of this term.}. This leads to the following data model for the residuals per transit:
\begin{equation}\label{eq:StatModel3}
    \begin{cases}
        \mathcal{H}_0: y_k=  \mu +\epsilon_k, \\
        \mathcal{H}_1: y_k = \mu + A \sin(2\pi f t_k + \varphi) + \epsilon_k,
    \end{cases}
\end{equation}
where $y_k:=~y(t_k)$,  $\epsilon_k\sim {\cal{N}}(0, \overline{\gamma}_k)$ with $ \overline{\gamma}_k$ is the std given in \eqref{gammabar} for a transit $k$. 
Finally,  for each target, the transit data  above are collected in a window composed of a number ($K$) of transit data points, most often 10 points. This leads to a time series ${\bf{y}}:=~[y_1,\cdots, y_K]^\top$. As discussed in L24, the detection method relies on a GLSP analysis that compares the sums of the weighted least square residuals obtained by fitting only a constant and a constant plus a sinusoid. The ``significance'' of the score reflecting this comparison is calibrated by MC simulations with a p-value.

Note that in the transit data model \eqref{eq:StatModel3} the systematic term $\mu$ is not indexed by $k$. This is an approximation, as this bias indeed slightly varies within the window. One way to account for such possible  variation is to model the bias signal as a low-order polynomial and to inject this in the GLSP analysis of the time series composed of the $K$ data points (App. \ref{app:gglsp}). While this makes the processing computationally more heavy (any additional computation has to be multiplied by about $10^5$ candidates and by $10^4$ MC simulations for each candidate), our investigations show that the results of such a model are often very similar to those using a constant $\mu$ in model \eqref{eq:StatModel3} (Fig.\ref{fig:sameGLSP}, top panel for a typical example). Consequently, we opted for this simpler model for almost all candidates, except those for which a trend was clearly detected (Sec.~\ref{sec:trendy}).

Figure \ref{fig:err_dr3_fpr} compares the distributions of the std per transit, for transit data computing using  1) a simple mean per transit of the observation residuals, and the corresponding std, being provided by \gdrthree data (filled gray); 2) a simple mean per transit of the observation residuals of \gfpr data, the std being computed as $(\sum_{i=1}^N \gamma_i^2)/N$ (red dotted); 3) the weighted mean of \gfpr data in  Eq.~\eqref{eq:StatModel2}, the std being computed by Eq.~\eqref{gammabar}  (solid line violet).  Note that the residuals per observation (CCD) were not available to us in  \gdrthree, making the computation of the weighted mean not possible. This figure shows that when computing the transit data as a simple mean, the resulting data are similar for \gdrthree and \gfpr. However, \gfpr data computed as the weighted mean as in Eq.~\eqref{eq:StatModel2} have a substantially lower dispersion. Indeed, samples affected by larger errors are  weighted less, which makes the final estimate more accurate.

\section{Confidence intervals}
\label{sec:ci_est}

Let us denote by $\mathcal{P}(f)$ the GLSP of the time series ${\bf{y}}:=~[y_1,\cdots,y_K]$. %
The wobble frequency $\widehat{f}$ is that of the largest peak in the periodogram: $\widehat{f}:=~ \arg\max_\nu \mathcal{P}(\nu)$.

The estimated amplitude $\widehat{A}$ and phase $\widehat{\varphi}$
are obtained through a standard weighted least squares (WLS) fit of a sinusoid plus constant to the data\footnote{The time series can be written as
$y_k=~\beta_0+\beta_1\cos(2\pi \widehat{f} t_k)+\beta_2\sin(2\pi \widehat{f} t_k)+\epsilon_k$ for $k=1,\cdots,K$. The WLS estimation leads to estimates $\widehat{\beta}_0$, $\widehat{\beta}_1$ and $\widehat{\beta}_2$ (App. \ref{app:gglsp}). The estimated parameters are then $\widehat{\mu}=~\widehat{\beta}_0$,   $\widehat{A}=~\sqrt{\widehat{\beta_1}^2+\widehat{\beta_2}^2}$ and $\widehat{\varphi}=~\mathrm{atan}(-\frac{\widehat{\beta}_2}{\widehat{\beta}_1})$.  }.
The question addressed now is to provide a reliable CI for these quantities.

\subsection{Algorithm}
\label{subsec:algo}
This algorithm is inspired by bootstrap ideas \cite{efron1979,efron_2010}.
%empirical, and results from a comparison of several methods not detailed here. Deriving  guarantees for bootstrap methods is difficult {\bf{[Ref]}} but 
We provide in this section numerical studies investigating the reliability of CI from this algorithm. The pseudo-code summarising the algorithm for CI estimation (Algo~\ref{algo:ci_fpr}) is presented in App.~\ref{app:algo}. The idea is to estimate the CI from an empirical distribution of the estimates of the amplitude and period through MC simulations. 

Starting from the estimated parameter of the wobble $\widehat{f}$ and $\widehat{A}$, and the other WO parameters, we generate $M$ simulated time series consisting of a sinusoid with the estimated wobble amplitude, sampled at the considered epoch $\bt$, to which are added a systematic offset $\mu^{(i)}$ and random noise $\bn^{(i)}$. In order to add diversity in the signal and thus robustness to the procedure, we generate, for each simulation $i$, a new phase (drawn from a uniform distribution) and a new systematic offset. This offset is consistent with \gaia residuals and modelled as the realisation of a Laplacian random variable with the parameters  (step 3 of Algo~\ref{algo:ci_fpr}, and for more details see App.~A of L24). The noise std at the transit level was derived in the previous section ($ \overline{\bgamma}$ in Eq.~\eqref{gammabar}). In practice, to be conservative, the values of the std in Algo~\ref{algo:ci_fpr} are taken slightly larger than those computed in Eq.~\eqref{gammabar} because we have seen in Sec.~\ref{subsec:prop_gaia} that those are slightly underestimated\footnote{Precisely, we use $\sigma := ~\sqrt{\tilde{\mu}^2 + \overline{\gamma}^2}$, where $\tilde{\mu}$ is the systematic error component in AL direction as estimated in L24; note, however, that the value of $\sigma$ is very close to $\overline{\gamma}$ because generally $\tilde{\mu}\ll \overline{\gamma} $.}.

\begin{figure}[htpb]
    \centering
   \includegraphics[width=\linewidth]{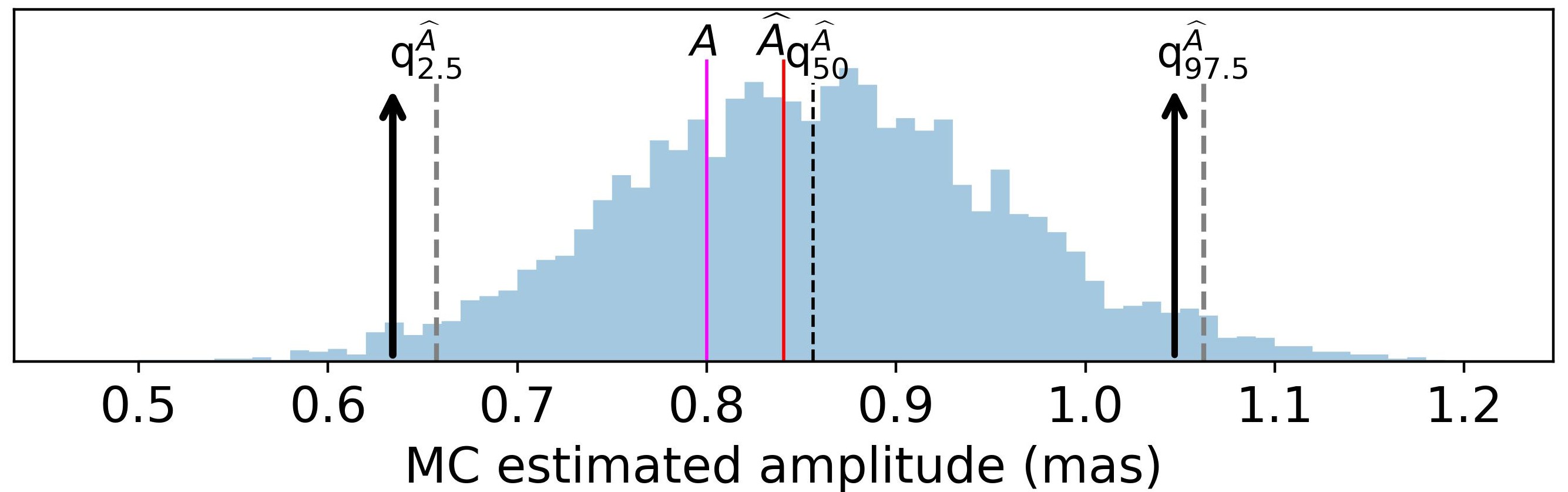}
    \caption{Illustration of the CI computation for the amplitude estimate $\widehat{A}^{(i)}$ (blue distribution) produced by Algo~\ref{algo:ci_fpr} showing the true (unknown) wobble amplitude $A$ (violet), the estimated amplitude $\widehat{A}$ (red), the median $q^{\widehat{A}}_{50}$ (dashed black), the quantiles $q^{\widehat{A}}_{2.5}$ and $q^{\widehat{A}}_{97.5}$ (dashed gray), and the final CI ${\mathcal{I}}_{{A}}$ for $\widehat{A}$ (black arrows).}
    \label{fig:CI_dist}
\end{figure}

For each Monte Carlo time series, frequency, period, amplitude and phase are re-estimated. This leads to a distribution of estimated frequencies (or periods) and of amplitudes, from the quantiles of which the CI can be computed. This procedure is illustrated in Fig.~\ref{fig:CI_dist} for a CI regarding the amplitude (${\mathcal{I}}_{{A}}$). The nominal (unknown) wobble amplitude is in violet ($A=~0.8$ mas). The estimated amplitude $\widehat{A}\approx 0.84$ mas is in red. The simulated data from which  $\widehat{A}$ was estimated were generated as a sinusoid with amplitude $A$ and random (uniform) phase, sampled at the same epochs $\bt$ as one (arbitrary) WO from a \gaia target, with values of the added offset and noise std in agreement with the \gfpr data model. Algorithm~\ref{algo:ci_fpr} produces the distribution of estimates $\widehat{A}^{(i)}$ shown in blue, whose median $q^{\widehat{A}}_{50}\approx 0.86$ mas (black dashed line). The location of the quantiles $q^{\widehat{A}}_{2.5}$ and $q^{\widehat{A}}_{97.5}$  of the distribution are shown as the thick grey dashed lines. Here  $q^{\widehat{A}\star}:=~\max \{q^{\widehat{A}}_{97.5}-q^{\widehat{A}}_{50},q^{\widehat{A}}_{50}-q^{\widehat{A}}_{2.5} \}= ~q^{\widehat{A}}_{97.5}-q^{\widehat{A}}_{50}  \approx 0.205$ mas.  The resulting CI ${\mathcal{I}}_{{A}}$  is shown by the two black arrows. In this case, the CI does contain the true amplitude value. As will be shown in the next section, this indeed happens with probability around $95\%$ for most amplitudes.

\subsection{Changes with respect to \cite{liberato2024} and performance evaluation}
 
Algorithm~\ref{algo:ci_fpr} presents several important changes with respect to the previous implementation used in L24. First, the noise added in step 4 was uniform (in an interval ranging from $0$ to the estimated std of the error in \gdrthree), which indeed caused an underestimation of the noise effect. Second, the amplitude was estimated using the difference between the maximum and the minimum of the sinusoid fitted to the data, instead of using the WLS coefficients $\widehat{\beta_1}$ and $\widehat{\beta}_2$ mentioned above, which provided less accurate amplitude estimates.  Third, the quantiles $q^A_{2.5}$ and $q^A_{97.5}$ were estimated using binned histograms of $\widehat{A}^{(i)}$ and  $\widehat{T}^{(i)}$, which created a slight but unnecessary dependence of the estimated quantiles on the bin width used. In contrast, the quantiles computed as in steps 12 to 14 are obtained by a consistent estimator \citep{david2004}.
 
We turn now to the evaluation of the actual false coverage rate (FCR) of the derived CI intervals. The FCR is the probability that $A\notin {\cal{I}}_{A}$ and is calculated by the fraction of simulated cases where the CI did not contain the initial value.
Our approach is based on MC simulations. For a given wobble (sinusoidal signal) with known period $T$, amplitude $A$ and phase $\varphi$,  we compute a set of $100$ simulated data (time series). For each such time series, we estimate the amplitude and the period,  run Algo~\ref{algo:ci_fpr} and check whether $A$ and $T$ belong or not to the claimed CI.
The epochs $\bt$ and noise std $\boldsymbol{\sigma}$ change for each MC simulation, they are drawn randomly from those of the WO in \gfpr, and the offset is generated as a Laplacian random variable (Step 5 of Algo~\ref{algo:ci_fpr}). We repeat this experiment by varying  the nominal amplitude of the signal $A$, between $0$ and $3$ mas.

 \begin{figure}[htpb]
     \centering
     \includegraphics[width=\linewidth]{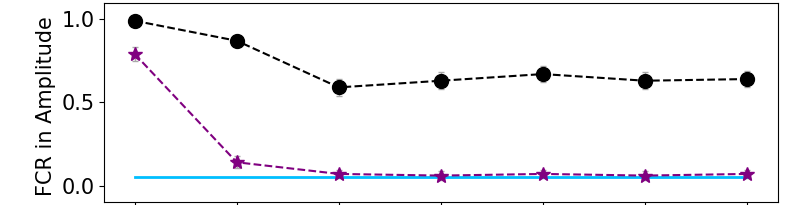}
     \includegraphics[width=\linewidth]{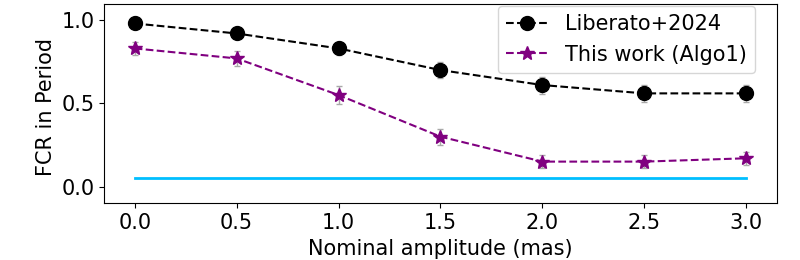}
     \caption{False coverage rate (FCR) with error bars (in grey) of the CI for the estimated amplitude (top) and the estimated period (bottom) for the two versions of the CI algorithms.  The horizontal solid blue line indicates the 95\% coverage targeted. }
     \label{fig:versions}
 \end{figure}

 In Fig.~\ref{fig:versions} we show the empirical FCR as a function of the initial (or nominal) value of the wobble amplitude. The black dots correspond to the algorithm used in L24 and the violet stars to Algo~\ref{algo:ci_fpr}. We notice in the top panel that the CIs from Algo~\ref{algo:ci_fpr} reach the 95\% confidence level expected for input wobble amplitudes larger than about $1$ mas (see App.~\ref{app:performance}). In contrast, the algorithm from L24 (without the changes above) often fails to cover the true amplitude. Turning to the period estimation (bottom panel), we see that the estimation is more difficult, with yet better performances for Algo~\ref{algo:ci_fpr}. Here, this algorithm provides CI for the periods that are valid with probability $90\%$ instead of $95\%$ for wobble amplitudes about $2$ mas or more.

%*******************************************************************************************************************************************************************************************************************************************************

\section{Trendy residuals}
\label{sec:trendy}
\begin{figure}[htpb]
     \centering
     \includegraphics[width=\linewidth]{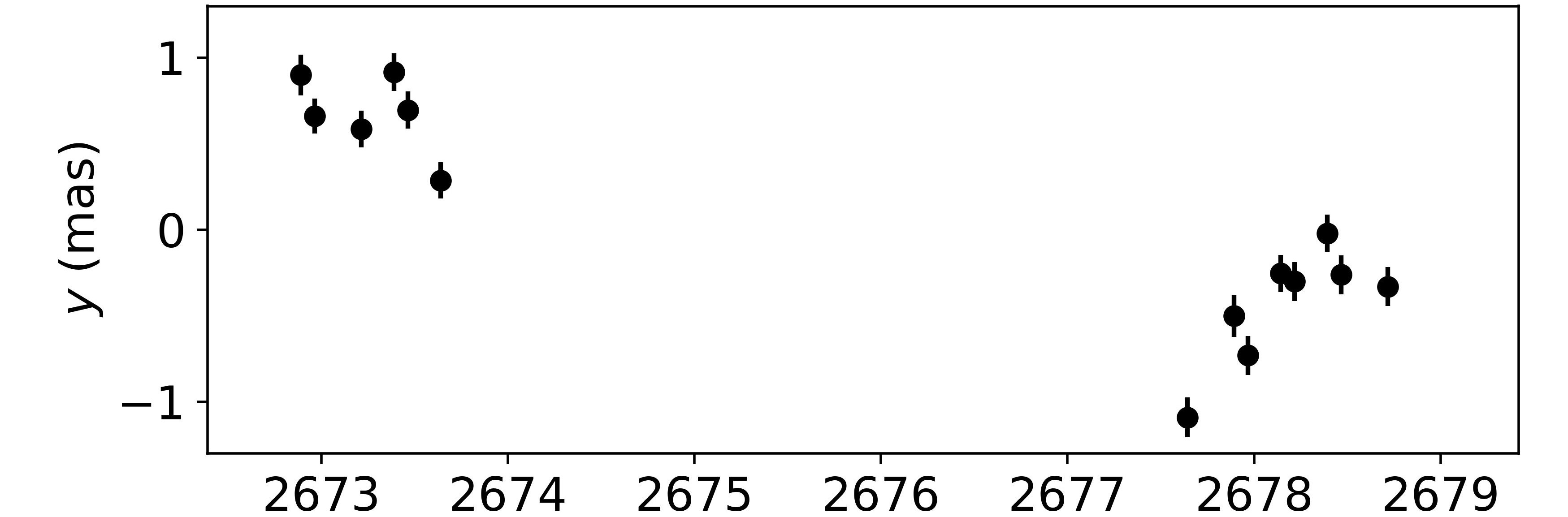}
     \includegraphics[width=\linewidth]{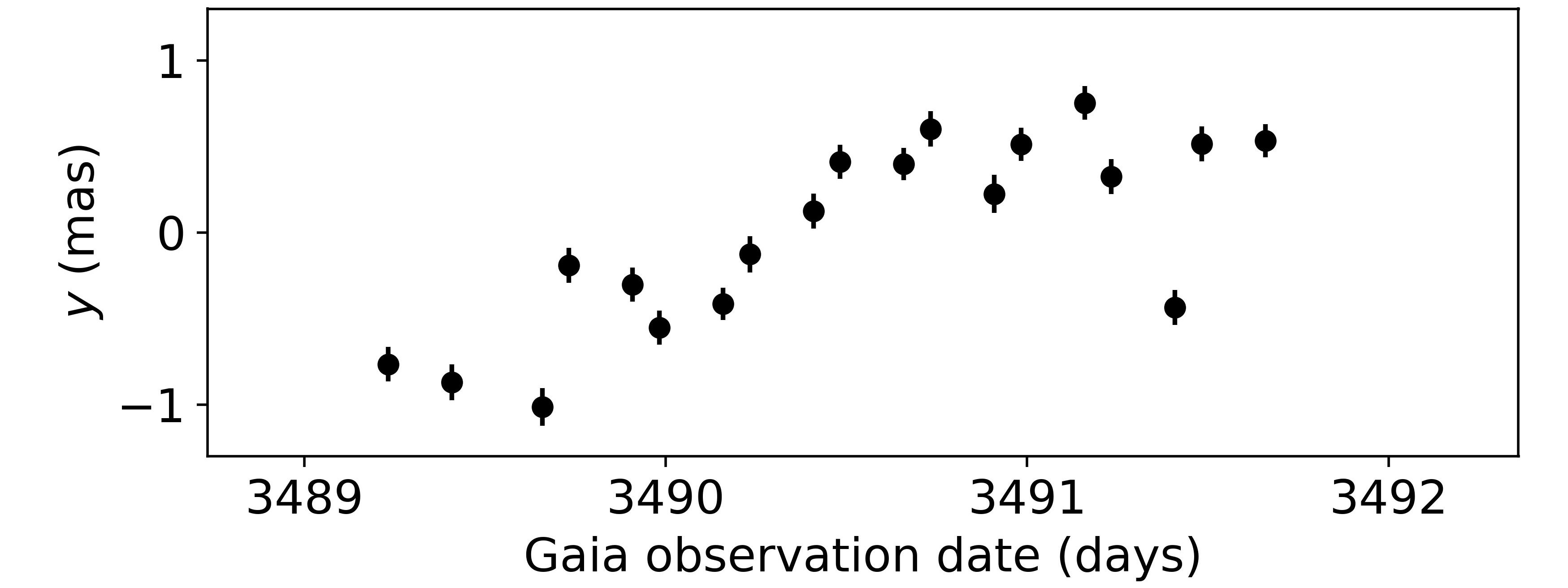}
     \caption{AL astrometric residuals per transit $y$ versus \gaia observation epochs for the known wide binary asteroid (317) Roxanne \citep{drummond2021orbit} in two different WOs. Both are flagged as ``trendy''.}
     \label{fig:roxanne}
 \end{figure}

For some (less than $2\%$) of the windows of observations explored, the residuals within show a global trend that dominates the time variation, see two examples in Fig.~\ref{fig:roxanne}. Such trends could be due to variation in the systematic offset discussed in Sec.\ref{subsec:model}, to other unknown artefacts, but also to wobble periods that are much longer than the WO. Hence, these cases deserve dedicated processing. To address this point, we set-up a two-step procedure: (1) A trend detection step, using a test on the Pearson correlation coefficient $c$ between epochs and residuals of the WO in question (App.~\ref{app:pearson}); (2) A dedicated period search using an extension of the GLSP that includes a linear trend in its data model (App.~\ref{app:gglsp}).

For step (1), we compute a p-value\footnote{We perform $10^4$ MC simulations where we estimate $c$ on simulated data sampled at the same epochs as the WO, using noise std and systematic as described in the data model. The p-value of $c$ corresponds to the proportion of this population of $10^4$ correlation coefficients (obtained with zero correlation between $\bt$ and $\by$) that is larger than $c$.} associated with each correlation score $c$. If this p-value is below $<0.5\%$ then the WO is flagged as a ``trendy WO''. For instance, for the two cases shown in Fig.~\ref{fig:roxanne}, the correlation coefficients are $c=~-0.87$ (top)  and  $c=~0.77$ (bottom), with both p-values less than $10^{-4}$.  At step (2), the trendy WO then go through the same procedure as the non-trendy ones, but using our implementation of the combined GLSP+trend instead of the conventional GLSP used in period search for the rest of the sample (App. \ref{app:gglsp}). For all of the WOs tested and a selection threshold of 0.5\%, one expects approximately 300 false positives under $\mathcal{H}_0$. We identify 1,433 trendy WOs, representing a significant excess relative to the null expectation. This excess indicates that a substantial fraction of the sample exhibits genuine correlated residuals. 

Adopting a conservative false discovery rate estimate, we infer that the majority ($\approx80$\%) of the selected objects are likely to correspond to real correlations rather than statistical fluctuations. However, among those, there are 207 WOs that correspond to the first or last observations available for such objects in \gfpr. Observations at the boundaries inherently contribute less new information in the orbital fitting procedure, leading to larger uncertainties and possibly affecting residuals near the start/end of the observation arc \citep{milani2010,spoto2018}. Therefore, as these 207 WOs, the trend detection presents a larger possibility of not being due to physical effects, we decided to omit them. Finally, among the remaining 1226 trendy WOs, there are 45 objects, including the known binary (317) Roxanne, with two WOs flagged as trendy. It means that the linear trends are less likely to be spurious for these objects, since they are detected in different epochs, which makes them interesting targets for further studies.

%*******************************************************************************************************************************************************************************************************************************************************

\section{Statistical selection of the candidates}
\label{sec:stat_sel}

In L24, the statistical selection of the binary candidates was done in two steps: (I) Compute a p-value associated with the WO using the \gaia noise model at hand and select the object if this p-value is less than $5\%$; and (II) Compute a ``quality factor'' $Q$ and select the candidate if $Q$ is larger than $50\%$, meaning that the detected period value should be found back (within some tolerance) with probability larger than $50\%$ in simulated data of a noisy sinusoid with that period.

Regarding step (I), the selection process for the p-values was a simple threshold rule. This has the advantage of providing an idea of the average number of false detections when $\mathcal{H}_0$ is true for all candidates (this number is $5\%$ of the total number of cases tested). However, a more interesting criterion can be the proportion of false discoveries (or false discovery rate, FDR) in the selected list.

The Benjamini–Hochberg (BH) procedure \citep{benjamini1995} is used for that purpose here. This method guarantees to control the FDR if the p-values in the sample are independent and uniform. In our case, independence of p-values holds from the independence of the data from one transit to another. The uniform distribution depends on the accuracy of the noise model\footnote{By definition, if we denote by $S$ the random variable corresponding to the score of the GLSP (i.e., the value of the highest peak in the periodogram), a p-value $p$ for a particular value $s$ (a realization of $S$) is $p:=~\mathrm{Pr}(S>s\;|\;{\mathcal{H}}_0)$. Hence, the probability that the random variable  $P$ is less than $p$ is Pr$(P<p \;|\;{\mathcal{H}}_0) =~\mathrm{Pr}(S>s\;|\;{\mathcal{H}}_0)=~s$ (since any score larger than $s$ will have p-value less than $p$) showing that $P$ is uniform. This also shows that uniformity of the p-values  holds as long as Pr($S>s$) is accurately estimated. If this calibration is not accurate, small p-values may be more likely than expected, leading to increased and (worse) uncontrolled false alarm rate.} and will also be shown to hold empirically in this section. 

To choose the target FDR to perform the selection in step (I), we must take into account several points: 1) Low p-values may arise from statistical flukes of noise, from genuine wobble signals, or in some more rare cases from other sources (e.g. systematic or instrumental effects, undetected long-term trends, locally underestimated uncertainties,...); 2) We wish the list to contain some known binaries (even if their signatures is weak in the data, with not so small p-values). 3) We can afford a list with a large proportion of false discoveries because the subsequent physical validation step (Sec.~\ref{sec:physical}) is expected to remove a large fraction of them. For these reasons, the adopted target FDR is $75\%$. 

Regarding step (II), we mentioned in Sec. 4.3 of L24 that while the $Q$ factor can be valuable information in some circumstances, it can also be misleading, as, conversely, large values of $Q$ may also arise from pure noise flukes, and true but weak detections may lead to low values of $Q$. Hence, in this work decided not to use the $Q$ factor in the selection, but it remains encapsulated in the provided CI interval. 

In the rest of this section, we first provide a new method for statistically combining the p-values of objects having more than one WO, and we apply the whole statistical detection pipeline  to \gfpr data. Finally, we present numerical  tests aimed at verifying the validity of the approach.

\subsection{The method applied to \gfpr}

From the $\approx$ $157,000$ objects in \gfpr, $47,896$ objects contain at least one WO within the 66 months of \gfpr data. It translates into $60,152$ WOs to be analysed. For this data set, the transit-averaged residuals from the orbital fit are computed as in Eq.~\eqref{eq:StatModel2} and the respective errors as in Eq.~\eqref{gammabar}. For each WO, we  apply a period analysis using GLSP, as briefly explained in Sec.~\ref{sec:intro} (App.~\ref{app:gglsp}), with nifty-ls implementation in Astropy \citep{vanderplas2018,Garrison_2024}, and estimate the corresponding empirical p-values through 10,000 MC simulations.

With the full sample of $60,152$ p-values from all of the WOs, we first separate the objects with a single WO in \gfpr from those with multiple WOs, and perform the BH-based selection independently in each group at a target FDR of 75\% (Tab.~\ref{tab:wo_dist}). For the $37,354$ objects with only one window, each object is associated with a single p-value, which is treated independently. As a result, we obtain \swBH selected objects, of which about 25\% ($\approx$~178) are expected to be true detections.

For the $10,542$ objects with multiple windows, an independent p-value is computed for each window. Performing the procedure considering both windows as a single data set requires a different approach. The method would need to take into account the variations in the geometry of the observation, which is beyond the scope of this work. Now, a fact needs to be considered: while eventually individual windows may yield marginal or no detections, their combined behaviour may increase the detection strength when considered jointly. Hence, our goal is to combine the multiple p-values associated with a given object in order to evaluate the global evidence against the null hypothesis. For that purpose, we apply to the multiple window objects two complementary methods of p-value combination.

The first is Fisher’s method \citep{fisher1948}, which accumulates evidence from the combined windows and is most powerful when many p-values are moderately small. It is therefore sensitive to weak but consistent signals spread over several WOs of an object. This method combines $n$ independent p-values, $p_1, p_2, \ldots, p_n$, by calculating:
    \begin{equation}
     k_F :=~ -2 \sum_{i=1}^{n} \ln(p_i).
    \end{equation}
    If the null hypothesis holds for all tests, $k_F$ follows a $\chi^2$ distribution with $2n$ degrees of freedom. From $k_F $, the Fisher combined score can be written as:
    \begin{equation}
        p_{\textrm{Fisher}}:=~1-\Phi_{\chi^2_{2n}}(k_F),
    \end{equation}
     with $\Phi_{\chi^2_{2n}}$ the CDF of a $\chi^2_{2n}$ random variable. This is again a p-value if the initial p-value sample is independent and uniform (Fig. \ref{fig:pvals_dist}), which is indeed our case.
        
The second is the min(p) method \citep{tippett1931}, which is sensitive to the presence of at least one strong signal. This method is powerful in scenarios where a strong detection appears in only one of the WOs of an object. The min(p) method takes the smallest value among them:
    \begin{equation}
    k_m :=~ \min\{ p_1, p_2, \ldots, p_n \}
    \end{equation}
    and adjusts it, accounting for the number of values in the set. The p-value associated with $k_m$ can be easily computed as :

    \begin{equation}
    p_{\mathrm{min(p)}} := ~1 - (1 - k_m)^n
    \end{equation}
    
By using min(p) and Fisher, we probe the two limiting and physically relevant cases—dominance by a single window versus coherent evidence from many windows—while relying on methods with simple, well-defined null distributions and a straightforward interpretation without the need for assumptions or previous knowledge on the true detections distribution \citep{heard2018,vovk2020}.

\begin{table}[htpb]
\caption{Distribution of WOs, objects and WOs per object in each group, along with the total number of WOs and objects in the analysed sample.} 
    
    \begin{tabular}{c c c c}
    
    \multirow{2}{1.5cm}{\# of WO per object} & \multirow{2}{*}{\# of WO} & \multirow{2}{*}{\# of objects} & \multirow{2}{2.5cm}{\# of objects selected with BH}\\
    & & &\\\hline
    1   & 37,354 & 37,354 & 713\\
    $>$1& 22,798 & 10,542  & 735\\\hline
    2 & 18,020 & 9,010 & 596\\
    3 & 4,107 &  1,369 & 129\\
    4 & 584 & 146 & 8\\
    5 & 75 & 15 & 2\\
    6 & 12 & 2 & 0\\\hline
   total & 60,152 & 47,896 & 1,448
    \end{tabular}
        \label{tab:wo_dist}
\end{table}

After applying both combination methods and using the BH selection with a target FDR of $75\%$ on the two sets of combined p-values, we obtain \mwBHminp objects selected from the min(p) method, and \mwBHfisher selected with the Fisher method. The union of the two multi-window selected object samples leads to a total of \mwBH objects selected, as shown in Tab.~\ref{tab:wo_dist}. Finally, by getting the union\footnote{Note that while the FDR is controlled at the target level on each set (single, min(p) and Fisher), this is not guaranteed theoretically for the union data set.} of the single-window objects selected with the multi-window objects selected, we obtain a final sample of 1,448 statistically selected binary candidates from \gfpr astrometric data.  

For the objects with multiple windows, all the WOs associated with a selected object are used in the subsequent selection steps (discussed in the next sections), even if only one of the WOs passes successfully the BH selection threshold (because a WO that does not present a signal strong enough to be selected with a low p-value, can still be useful to confirm the period detected in the main WO).

\subsection{Performance evaluation using control simulations}

In order to assess the reliability of the statistical selection procedure and to quantify the level of spurious detections expected in the absence of any true astrometric signal, we perform simulations on signal-free data. The goal is to verify that the adopted period search, p-value estimation, combination procedures, and BH selection behave as expected under the null hypothesis, and to provide a reference against which the results obtained on the real \gfpr data can be interpreted.

1.)\textit{ FDR control}.
 We performed 10,000 simulations on sets of 60,000 random p-values mimicking the single and multi-window samples, applying the same p-value combination procedures, and we obtained an average final FDR of 74.5\% for the non-combined sample, 74.9\% for the Fisher-combined data set and 74.6\% for the min(p)-combined p-values, confirming that our procedures guarantee the expected FDR in all cases. 

2.) \textit{Full statistical detection pipeline}.
To compare our results on the \gfpr data set with those obtained from mirror but signal-free data sets, we performed 10 independent simulation runs. For each simulation run, we used all of the $\approx$ 48k objects explored in the candidates search, but replacing the residuals by noise plus systematic offset as described in the data model (Sec.\ref{sec:errors}). We then submitted each simulated data set to the exact same pipeline used for the \gfpr exploration. With this approach, we can ensure that the only (or most important) difference between the simulations and the \gfpr is the certainty that no wobble is present in the simulations.

 \begin{figure}[htpb]
     \centering
    \includegraphics[width=1.03\linewidth]{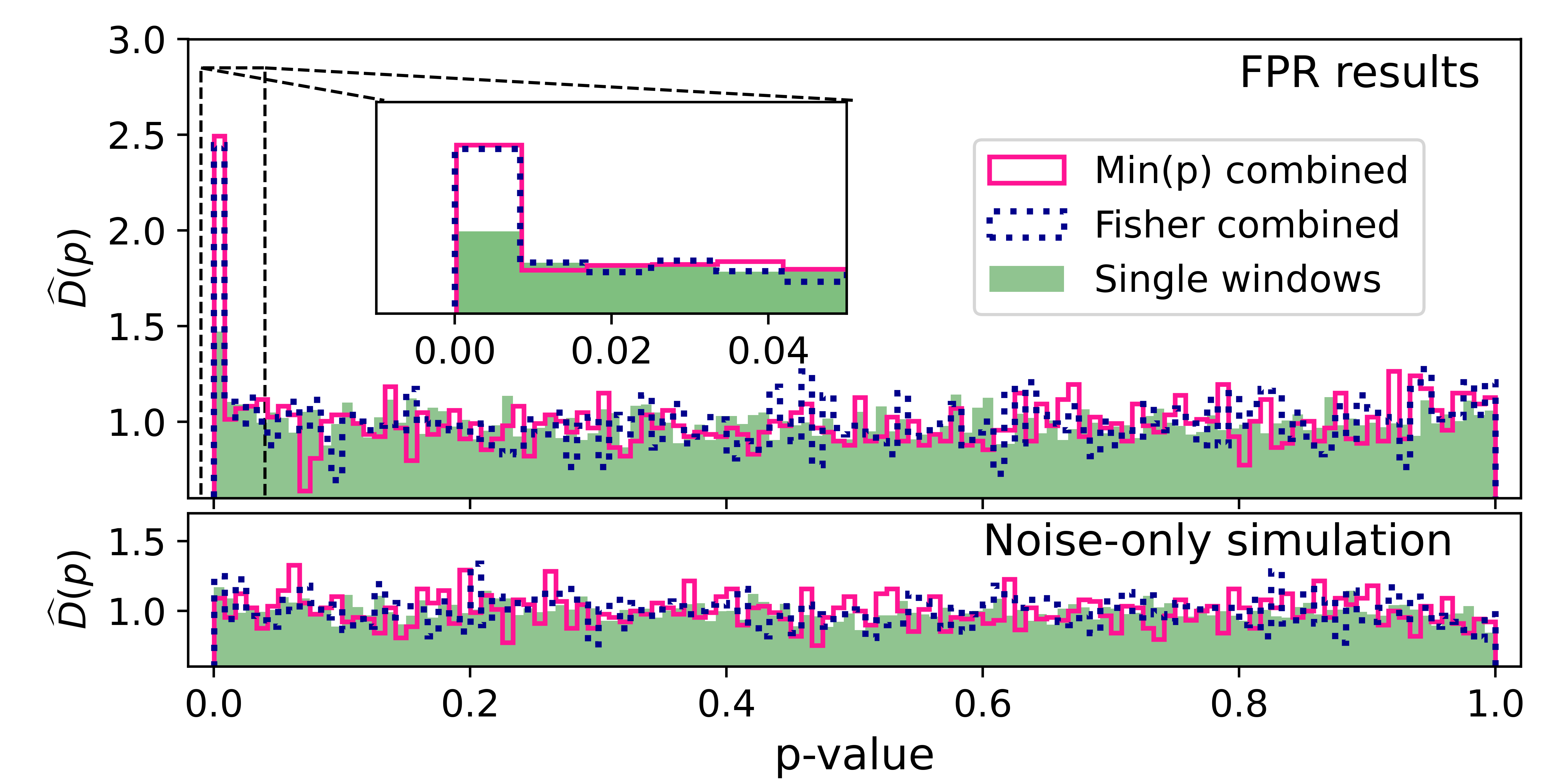}
     \caption{Estimated density distribution $\widehat{D}(p)$ of p-values for the single window objects (green), multiple window objects with p-values combined through the Fisher method (magenta) and combined with the min(p) method (blue). The bottom panel plot shows the results from one noise-only simulation run, while the top panel shows the p-value distributions from the FPR data. The inset panel is a zoom on the area delimited by the dashed line.}
     \label{fig:pvals_dist}
 \end{figure}

A first interesting result regards trend detection. We detected about 300 trendy WOs per simulation run, as expected due to the 0.5\% threshold adopted. This shows that it is unlikely that most of the 1226 WOs in which we detected trends (Sec.~\ref{sec:trendy}) are caused by noise fluctuations, especially for the 45 objects with two trendy WOs. Therefore, they deserve some further investigation. A second result concerns the distribution of p-values. Under ${\mathcal{H}}_0$, p-values are expected to follow a uniform distribution on [0,1], since they are derived from the quantiles of the test statistic distribution under noise-only simulations. As discussed before, this uniformity depends on the accuracy of the adopted noise model and its parameters: any mismatch leads to an incorrect calibration of the quantiles and hence to departures from uniformity, for instance in the presence of genuine signals or systematic effects.

Figure~\ref{fig:pvals_dist} illustrates this behaviour. The bottom panel shows the empirical p-value distributions (for the single, Fisher, and min(p) statistics) obtained from one of the 10 control simulation runs. Those are consistent with a uniform distribution, as expected, because in the simulations, the noise is generated according to the model, so the p-values are well  calibrated. The top panel shows the corresponding distributions for the \gfpr data. Two features are apparent: first, a broad plateau, indicating that for most objects the p-values are consistent with uniformity and that the noise model used to calibrate the GLSP scores (Sec.~\ref{sec:errors}) is appropriate; second, a clear probability excess at the smallest p-values. This excess indicates a subset of objects whose behaviour is inconsistent with ${\mathcal{H}}_0$, likely due to binary systems and other effects not consistent with the noise model.

\begin{table}[htpb]
\centering
\caption{BH selection counts from simulations and the \gfpr asteroids' data.}
\begin{tabular}{l|c|ccc|c}
Run & Single & Fisher & min(p) & union & Total \\\hline
sim 1  & 53  & 11 & 87 & 89 & 142 \\
sim 2  & 26  & 11 & 70 & 71 & 97  \\
sim 3  & 11  & 9  & 6  & 11 & 22  \\
sim 4  & 14  & 56 & 48 & 76 & 90  \\
sim 5  & 14  & 62 & 79 & 100 & 114 \\
sim 6  & 74  & 95 & 41 & 98 & 172 \\
sim 7  & 54  & 24 & 30 & 41 & 95  \\
sim 8  & 31  & 24 & 12 & 24 & 55  \\
sim 9  & 94  & 12 & 2  & 13 & 107 \\
sim 10 & 140 & 35 & 9  & 35 & 175 \\
\hline
\rowcolor{myblue} sim mean & 51.1 & 33.9 & 38.4 & 55.8 & 106.9 \\
\rowcolor{myviolet}
\shortstack[l]{\gfpr} & 713 & 588 & 605 & 735 & 1448\\
\end{tabular}
\label{tab:sims_sel}
\tablefoot{The second column indicates the amount of objects selected from single-windows' list; the third and fourth columns show the amount of objects selected from the multi-windows objects with p-values combined through Fisher and min(p) methods, respectively; the fifth column shows the number of objects resulting from the union of Fisher and min(p) selected objects; and the last column shows the total amount of objects selected in each run.}
\end{table}

Table~\ref{tab:sims_sel} summarises the results from the control simulations and from the \gfpr data exploitation. We can see that the number of WOs selected by our procedure on the control simulations is one order of magnitude smaller than those obtained in the \gfpr period search. These results support the idea that, although some of these detections can be spurious (due to unknown artefacts or modelling errors), there should be a significant number of real period detections in our results from \gfpr residuals exploration. In fact, in the absence of any such effect, roughly $25\%\; (\approx350)$ should be true detections.

%*******************************************************************************************************************************************************************************************************************************************************
\section{Physical validation of candidates}
\label{sec:physical}

After assessing the statistical relevance of the periods detected in the astrometric residuals, we check if the signal detected in the astrometric data is consistent with plausible physical parameters for a binary system. 

As explained in L24, with the approach adopted, we derive the minimum bulk density profile as a function of the mass ratio by combining the simple binary wobble model from \cite{hestroffer2010gaia} with Kepler's third law. The resulting expression combines the measured parameters (period and amplitude of the signal) and data that we extract from literature (diameter of the equivalent sphere) using the SsODNet database \citep{berthier2022ssodnet}. Setting thresholds on plausible densities \citep{carry2012density, scheeres2015asteroid} allows us to determine possible ranges of size ratio and separations. By adopting a density range 0.8--5.5 $g/cm^3$, we obtain a list of 988 WOs (729 objects) with values of minimum densities that fall within the chosen range.

Whenever necessary, we re-constrain (trim) the intervals of possible separations in one or both extremes. At minimum, we adopt the fluid Roche limit of the system \footnote{Asteroids in the size range of our targets are likely rubble piles, so adopting the fluid Roche limit is a more conservative approach.}. For the maximum separation, we set a limit of 20\% of the Hill radius\footnote{Usually the asteroid satellites are in compact configurations with separations $\sim$1\% of the Hill radius, so assuming the separation of our candidates up to 20\% is rather generous but still physically realistic.}, assuming that the density could be as high as 5.5~g/cm$^3$. 

We decided to reject candidates whose intervals of separation had to be re-constrained both in minimum and maximum. As their properties can be considered to be very weakly constrained, we consider that they are less reliable. Additionally, when this procedure rejects all WOs of multiple window candidates except one, we do not discard it only if its p-value is smaller than 4.2\%, the highest selected from the BH method in the multiple windows approach. After having applied these criteria, 353 binary candidates (and 421 WOs) remain.

We recall that, due to the single-dimensional nature of our approach, the wobble signature measured is a projection of the real photocentre offset, and is thus taken as a minimum value. Therefore, the derived binary parameters (density, separation, and mass ratio) are based on a minimum wobble amplitude, and therefore correspond to lower limits (or interval estimates) of the true values. Our estimates are limited by the signal measured in the astrometric residuals and may differ from values obtained with other observational techniques. Additionally, phase and shape effects (not accounted for in our simplified point-source model) can also introduce photocentre offsets comparable to or larger than the expected binary-induced signal in some configurations \citep{pravec2012small}. These effects may bias the inferred parameters or reduce the satellite wobble detectability, but modelling them requires prior knowledge of the system or a probabilistic approach, which is beyond the scope of this work.

%*******************************************************************************************************************************************************************************************************************************************************

\section{Results and discussion}
\label{sec:results}

\gaia is a complex system, whose scanning law, combined with the motion of the asteroids and the satellite, can potentially inject spurious frequencies in the data. It is then interesting to compare the distribution of the periods in our candidate sample to periods that arise in simulations with random noise.

\begin{figure}[htpb]
     \centering
     \includegraphics[width=\linewidth]{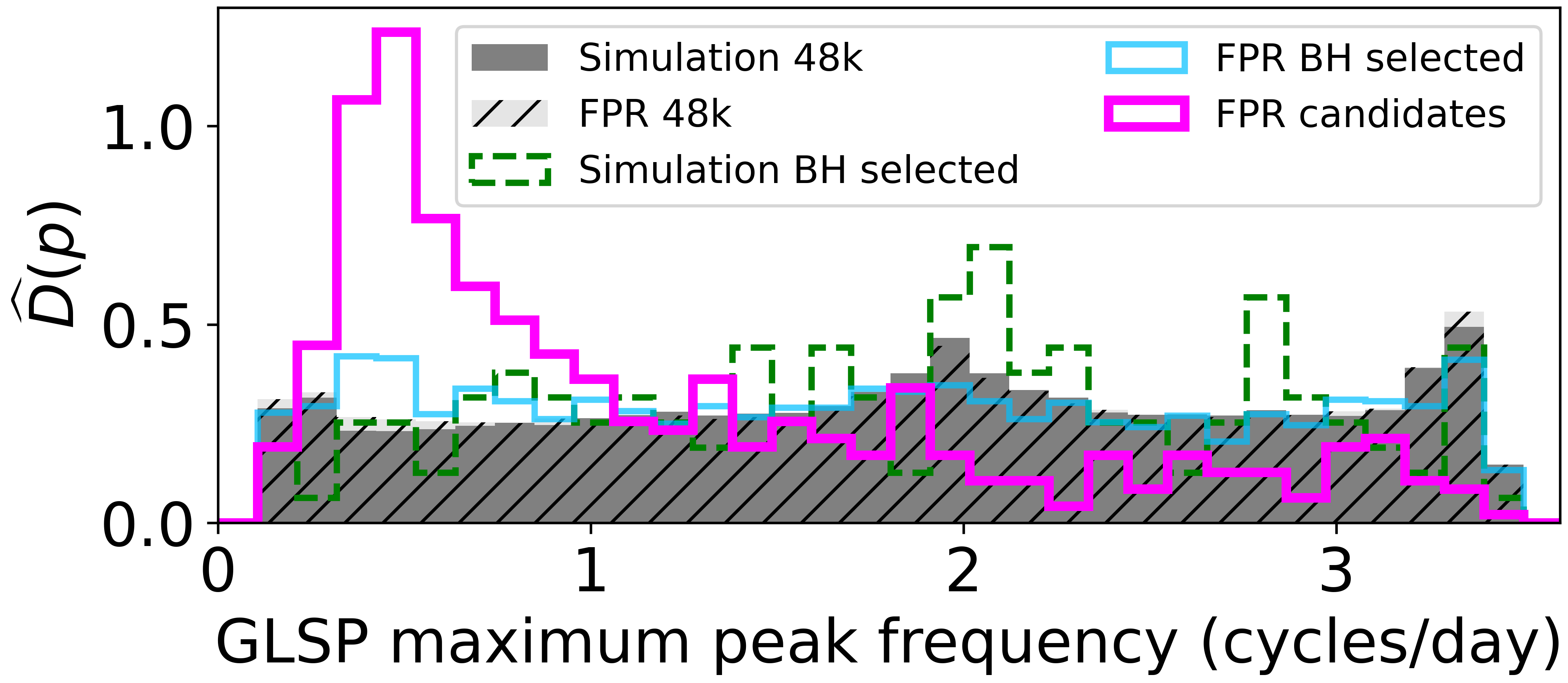}
     \caption{Density estimation on the distribution of the GLSP maximum peak frequencies for the results from one of the simulations explained in Sec.~\ref{sec:stat_sel} (gray solid bars); for objects selected by the BH method from the simulation (hatched bars); for all of the 48k objects from \gfpr (dashed green); for the \gfpr selected statistically with the BH method (thin blue); and the final list of \gfpr selected candidates (thick magenta).}
     \label{fig:freq_hist}
 \end{figure}

In Fig.\ref{fig:freq_hist} we see that the distributions from the full sample with all of the 48k objects in the simulation (grey solid bars) and in the \gfpr (dashed green) are very similar, indicating that the simulations successfully reproduce the dominant noise-driven behaviour of \gaia astrometric residuals. Additionally, we notice two bumps at frequencies around 3.6 and 2 cycles a day in both full sample distributions, which are probably aliases of the frequencies associated with the motions of the \gaia satellite \citep{cellino2024asteroid}.

The WOs statistically selected by the BH procedure, both in \gfpr and in the simulations, follow closely the same distribution as the full samples. This is consistent with the fact that we adopted an FDR at 75\%, implying that 3/4 of the selected objects at this stage are most likely false detections. However, remarkably, our final \gfpr selected sample shows a very different frequency distribution, suggesting that the physical validation step has probably eliminated the majority of spurious detections.

\begin{figure}[htpb]
    \centering
    \includegraphics[width=0.95\linewidth]{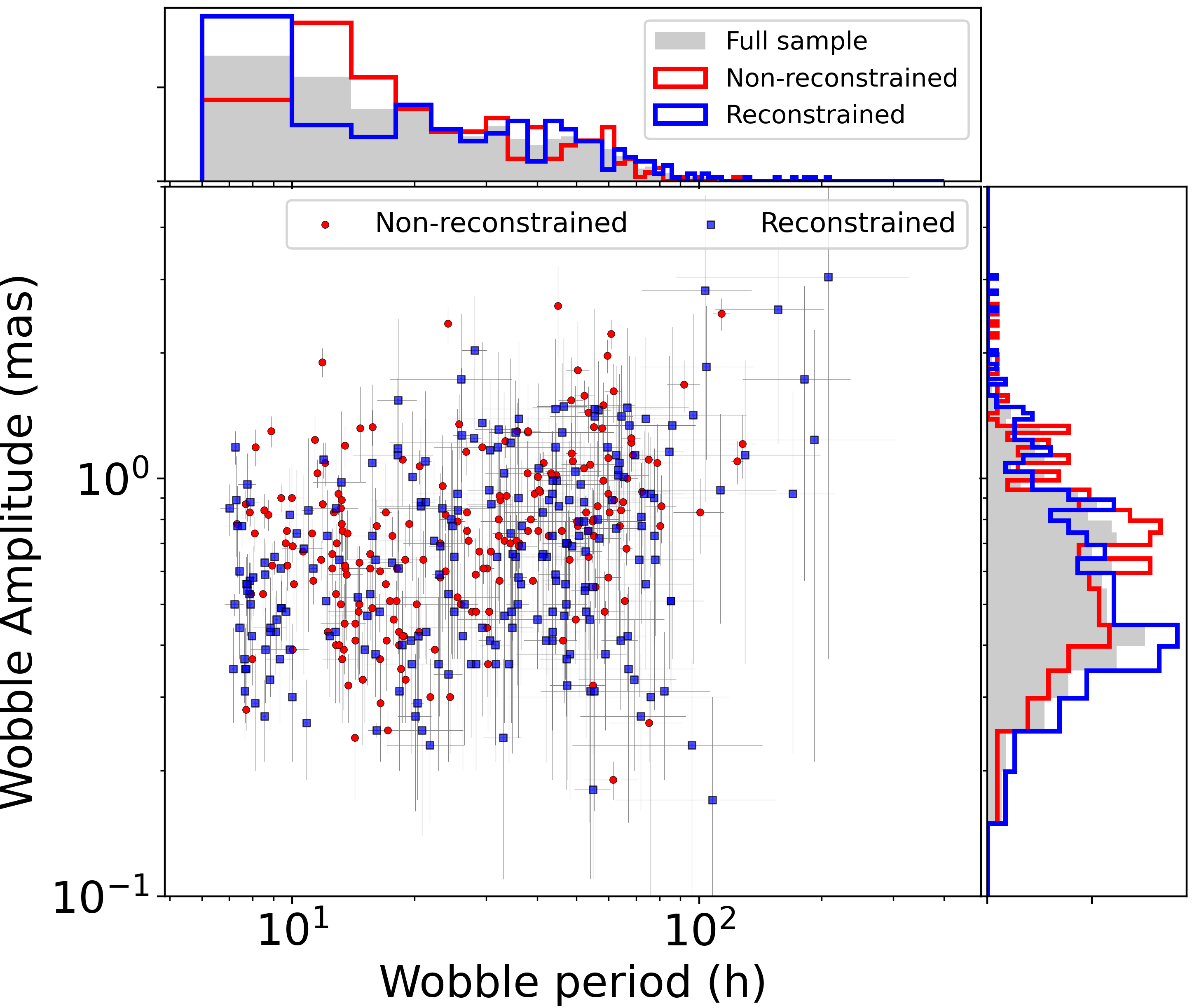}
    \caption{Amplitudes for the selected sample as a function of period. The pink dots correspond to separation intervals without the need for being re-constrained, while the blue squares were re-constrained, as explained in Sec.~\ref{sec:physical}. The histograms show the distribution for the entire sample (solid grey), while the coloured lines represent the same two categories as above.}
    \label{fig:wob_wob}
\end{figure}

Figure \ref{fig:wob_wob} shows the distribution of amplitudes and periods of the selected WOs. We can see that shorter periods ($<$24h) seem to be favoured by our method, as well as wobbles amplitudes 0.3--1~mas, compatible with those obtained from L24 (Sec.~\ref{subsec:compare_dr3}). The updates in our selection procedure do not affect the overall distribution much. However, these low values of amplitude tend to be poorly estimated, as we have shown in Fig. \ref{sec:ci_est}.

Among the final candidates, 27 objects exhibit at least one trendy WO, demonstrating that the detrending procedure (Sec.~\ref{sec:trendy}) effectively recovers signals for them. In addition, 45 objects show two WOs dominated by trends rather than fluctuations. A notable example is (317) Roxanne (Fig.~\ref{fig:roxanne}), a known wide binary \citep{drummond2021orbit} with a secondary orbital period ($\sim$12 days) much longer than the maximum WO span allowed by our selection. This indicates that, even when the wobble period cannot be directly measured, linear trends in astrometric residuals may signal the presence of wide binaries. For this reason, we select these 45 objects \citep[List available in][]{liberato2026zenodo} as targets deserving of deeper investigations.

\subsection{The known binaries selected}
\label{sec:kbins}
Cross-matching our candidates with the Johnston's Archive list of asteroids with satellites \citep{JArchive}, we find that among the 353 objects there are 9 known (or highly suspected) binaries that comply with all the selection parameters adopted. 
\begin{itemize}
\item (720) Bohlinia was identified in L24 and later confirmed by photometry \citep{gorshanov2025}. The period $T$=~17.418$\pm$0.006h (or double) from the photometry is consistent with our estimates $\widehat{T}$=~17.489$\pm$0.130h (L24) and $\widehat{T}$=~17.748$\pm$0.160h (this work). The separation estimate from the photometry $sep$=~73.47$\pm$0.01 km is close to our findings of 77.04$\pm$8.55 km in L24 and to the lower end of 87.18 km (this work). Differences can probably be attributed to the strong dependence on assumed physical properties, but especially to the new data model adopted, which led to slightly different wobble amplitude measurements. 

\item (1509) Esclangona is a known wide binary with size ratio $k\sim0.3$, $sep$ $\sim$140 km, and an $T$ $\sim$23~d \citep{merline2003esclangona}, well beyond the maximum observing window. We detect a trend with p-value$_{corr}=~0.541\%$, slightly above  our threshold. The period is close to the upper limit of the searched range. This case illustrates the limitation of our approach for wide systems, while still hinting at binarity through the trend in the residuals.

\item (1770) Schlesinger is a suspected binary based on reported mutual events, without reliable parameters. We select two WOs: one weak, and one strong, with $\widehat{T}$=~53.37$\pm$0.07h and p-value $<0.001\%$. Combined with previous suspicions, this makes Schlesinger a highly probable binary.

\item (1879) Broederstroom hosts a satellite with a reported $T$=~47.83$\pm$0.02h and $k=~0.34\pm0.02$ \citep{Benishek2024a}. We correctly detect $\widehat{T}$=~50.04$\pm$2.83h and estimate a smaller size ratio $0.12\le k_1\le0.24$. This discrepancy is most likely due to the observation geometry as explained in Sec. \ref{sec:physical}, where the projection of the wobble in the AL direction leads to a smaller amplitude and, consequently, mismatched mass ratio and separation estimations, similar to the case of (4337) Arecibo (see below).

\item (1967) Menzel  was identified as a binary candidate in L24 and confirmed  by photometry \citep{monteiro2024}. %,monteiro2026 
Photometry yields $T$=~63 h. We estimate $\widehat{T}$=~32.4308$\pm$1.01 h, about half, certainly due to the limitation in the length of the WO. The detection is very clear in the astrometry, both in \gdrthree and in \gfpr.

\item (2871) Schober is a known binary, with $T$=~42.47$\pm$0.02h and $k>0.28$ \citep{Benishek2023}. We obtain $\widehat{T}$=~50.65$\pm$21.65h and size ratio intervals of $0.1\le k_1\le0.3$ and $0.76\le k_2\le1$, consistent with the published parameters within uncertainties.

\item (4337) Arecibo is a synchronous binary with $k\sim0.19$ and $T\sim$32.97h \citep{gault2022new,tanga2023,liu2024}. We measure $\widehat{T}$=~35.4$\pm$4.0h, consistently, but estimate a maximum $k$ of $\sim$0.13. This underestimation is either due to unfavourable observation geometry (the AL-projected wobble amplitude reaching only $\sim$8.5\% of the system separation), or to the flattening of the components as mentioned in \citet{tanga2023}. 

\item (31450) Stevepreston has a satellite with $k>0.22$ and $T$=~53.47$\pm$0.07h \citep{Pray2015}. We find a $\widehat{T}$=~61.7$\pm$3.1h, $0.1\le k_1\le0.126$ and $0.96\le k_2\le0.98$. The period agreement is poor, still, we obtain a strong and clear detection with p-value $<0.001$. A possible explanation is the detection of a second, further (and maybe smaller) undetected satellite.

\item (55637) Uni is a $\sim$660 km KBO with a satellite at $4770\pm40$~km and $T$=~199.42~h \citep{brown2006,brown2013}, far longer than the \gaia consecutive observations span. We detect $\widehat{T}$=~45.02$\pm$1.23h, close to three times the primary rotation period ($\sim$14.4h), which could be the tracing of the photocentre shifts due to rotation of the primary. However, a more likely explanation would be the detection of a second satellite. A weak trend is still visible in the data, consistent with the presence of the known satellite.

\end{itemize}

\subsection{Are they all likely binaries?}
\label{subsec:likely_bins}
The motion of the photocentre with respect to the centre of mass of the system, showing up in the residuals of \gaia astrometry, can originate not only from satellites, but also from irregular shapes of single objects, provided that its amplitude is high enough to be detected \citep{kaasalainen2004,dell2012}. (21) Lutetia provides an example of this ambiguity, illustrated in \citet{tanga2023}. So, are all our current \gfpr candidates most likely binaries?
\begin{figure}[htpb]
    \centering
    \includegraphics[width=\linewidth]{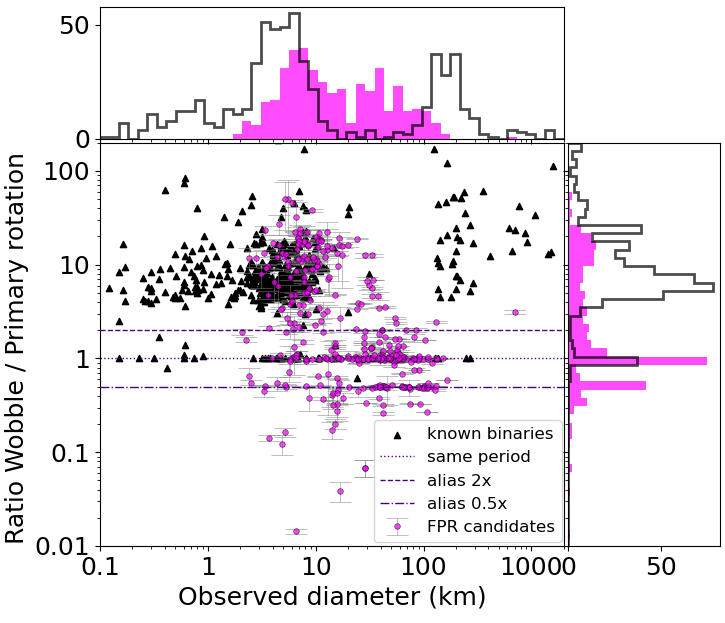}
    \caption{The population of known binary asteroids (black triangles) in comparison with our \gfpr binary candidates (pink circles), on the plane defined by diameter and wobble period normalised to the rotation period of the primary (as usually derived from photometry. A few peculiar objects with a very low period ratio appear at the bottom of the plot. They correspond to very slow rotators that could also have wrong photometric periods.}
    \label{fig:cand_dist}
\end{figure}

\begin{figure}[htpb]
    \centering
    \includegraphics[width=\linewidth]{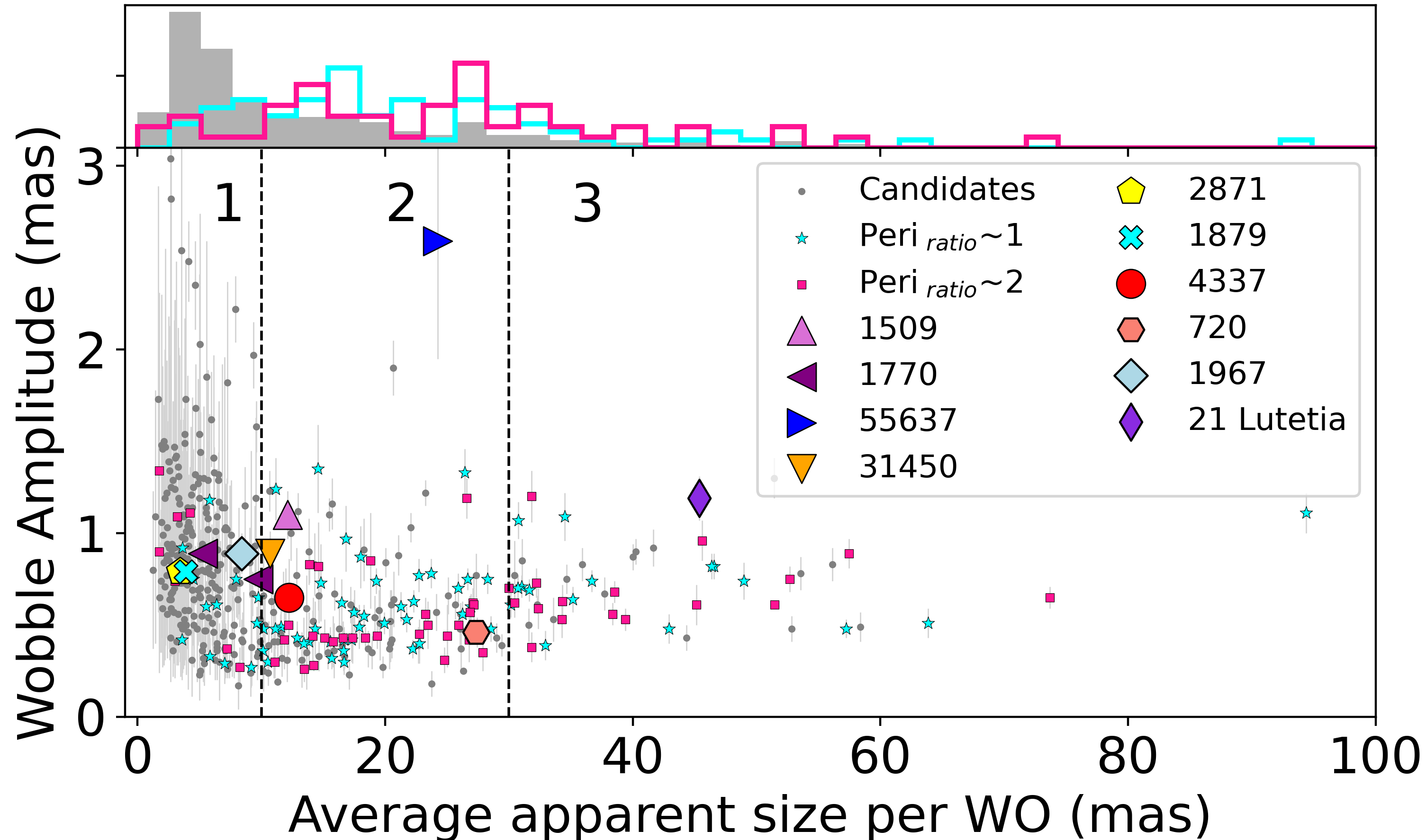}

    \caption{Distribution of \gfpr candidates' WOs in apparent size during the observation versus the wobble amplitude measured. The grey dots represent all the \gfpr WOs selected in the candidates' sample. The green stars represent the WOs in which the wobble period estimated has a period ratio of about $\sim$1 with respect to the rotation period observed for the object, while the pink squares are the WOs with period ratio$\sim$2.}
    \label{fig:res_windows}
\end{figure}

Figure~\ref{fig:res_windows} compares the wobble amplitude to the average apparent size (computed at each WO). It is divided into three regions. Region (1) contains objects with apparent sizes up to about 12 mas, corresponding to typical diameters smaller than 12--18~km in the Main Belt. Most of the known binaries selected in our sample lie in this region. They span a wide range of wobble amplitudes, mostly below 1.5 mas with larger uncertainties, consistent with a lower astrometric accuracy for fainter objects. The indicated known binaries in this region were discovered by photometry. This suggests that many candidates in region (1) may also be appropriate targets for this technique, as demonstrated for (1967) Menzel \citep{monteiro2024}. 

Region (3) contains the largest apparent sizes exceeding 30~mas, with typical diameters larger than 40--50~km. These objects exhibit a different behaviour, with small wobble amplitudes, despite the larger size than those in (1). The slightly increasing trend is suggestive of a wobble proportional to size, as expected for large, non-binary bodies. This size range hosts the largest fraction of wobble periods coinciding with the rotation of the primary (sometimes with an alias of double/half the value) as shown in Fig.~\ref{fig:cand_dist}.

The presence of (21) Lutetia in this sample is illustrative: we estimate a period (8.12 $\pm$ 0.03 h), closely matching its rotation \citep[8.168 h;][]{carry2010physical,sierks2011images}. Moreover, the Rosetta mission excluded the presence of satellites capable of producing the observed wobble \citep{bertini2012}. As the evidence is clear for this specific object, we exclude it from our candidate list. We keep the other candidates anyway and flag them in the list, to allow observers to make further verifications.

Finally, in the intermediate region (2), roughly corresponding to objects of 15--50~km, we find several confirmed binaries selected in our candidate sample. Wobble amplitudes here have moderate uncertainties, indicating stronger and cleaner signals. Most suspected synchronous binaries fall in this region, as shown by the histogram on the top of Fig.~\ref{fig:res_windows}, which corresponds to the underrepresented population of intermediate--size binaries consistent with formation via moderately catastrophic impacts \citep{durda2004}. Therefore, the fact that several known binaries detected are within these limits, plus most of the objects in this region are suspected synchronous binaries and are within the expected size range for such a formation mechanism, is another evidence that this is likely a binary--rich region.

Visual inspection of several DAMIT shape models \citep{DAMIT2010} reveals that confirmed binaries often have problematic single-body shape solutions: (4337) Arecibo has sharp edges; (720) Bohlinia appears unrealistically elongated, clearly a result of binarity in both cases \citep{durech2003}. Among our candidates with diameters $>$20 km, some shapes are smooth and spheroidal, but others present similar features. In region (2), (1105) Fragaria exhibits a sharp edge similar to Arecibo; (1127) Mimi (see also binary features in \citet{lallemand2025sf2a%,lallemand2026
}), (519) Sylvania, (542) Susanna, and (6475) Refugium display a significantly large elongation. In region (3), (303) Josephina shows similarities to Arecibo, whereas objects like (103) Hera, (538) Friederike, (605) Juvisia, (977) Philippa, (2906) Caltech, and (625) Xenia, with large flat surfaces, could reflect poorly modelled concavities instead of companions. 

In synthesis, candidates in regions (1) and (2) are the most robust in the sample, but we cannot exclude that at least some of the objects in region (3) have satellites, possibly with properties different from those of smaller objects. So, we decided not to discard them, as they can be interesting targets for other techniques. The users of our list should, anyway, remember the possible ambiguity of their cases. 

\subsection{Comparison with \gdrthree results}
\label{subsec:compare_dr3}
\gdrthree and \gfpr published astrometry for the same number of asteroids (about 150,000), obtained over 34 and 66 months, respectively. Correspondingly, the number of WOs that we extract increases from 30,030 to 47,896, about 60\% more. However, the lower number of binary candidates we present in this work (343, to be compared to 358 in L24) shows that the improvements we implemented led to a more conservative approach. In particular, instead of a simple threshold in \gdrthree, we used here a FDR-based selection rule, tuned so that on average roughly one fourth of selected objects ($\approx 1448/4=~362$) are not spurious detections. Interestingly,  $343$ candidates remain after the physical validation steps. Besides, we find that there is an overlap of 99 candidates ($\sim$28\%) selected in both \gdrthree and \gfpr. 

Among the objects selected in L24, and not in this work, are the known binaries (3220) Murayama, (5817) Robertfrazer, and (18301) Konyukhov. For Murayama and Robertfrazer, although the detected periods and amplitudes remain consistent between FPR and DR3, the modified data set led to higher p-values in this work, which reduced the statistical significance of the detections. As a result, they are not selected by the BH procedure. In the case of Konyukhov, the derived physical parameters do not satisfy the updated physical constraints, and the object is therefore excluded from the final sample. This exemplifies the fact that some candidates are close to the detection limits and different approaches can affect their detection, but the about 30\% overlap between the lists in L24 and in this work also demonstrates some robustness for stronger detections.

\begin{figure}[htpb]
    \centering
    
    \includegraphics[width=0.95\linewidth]{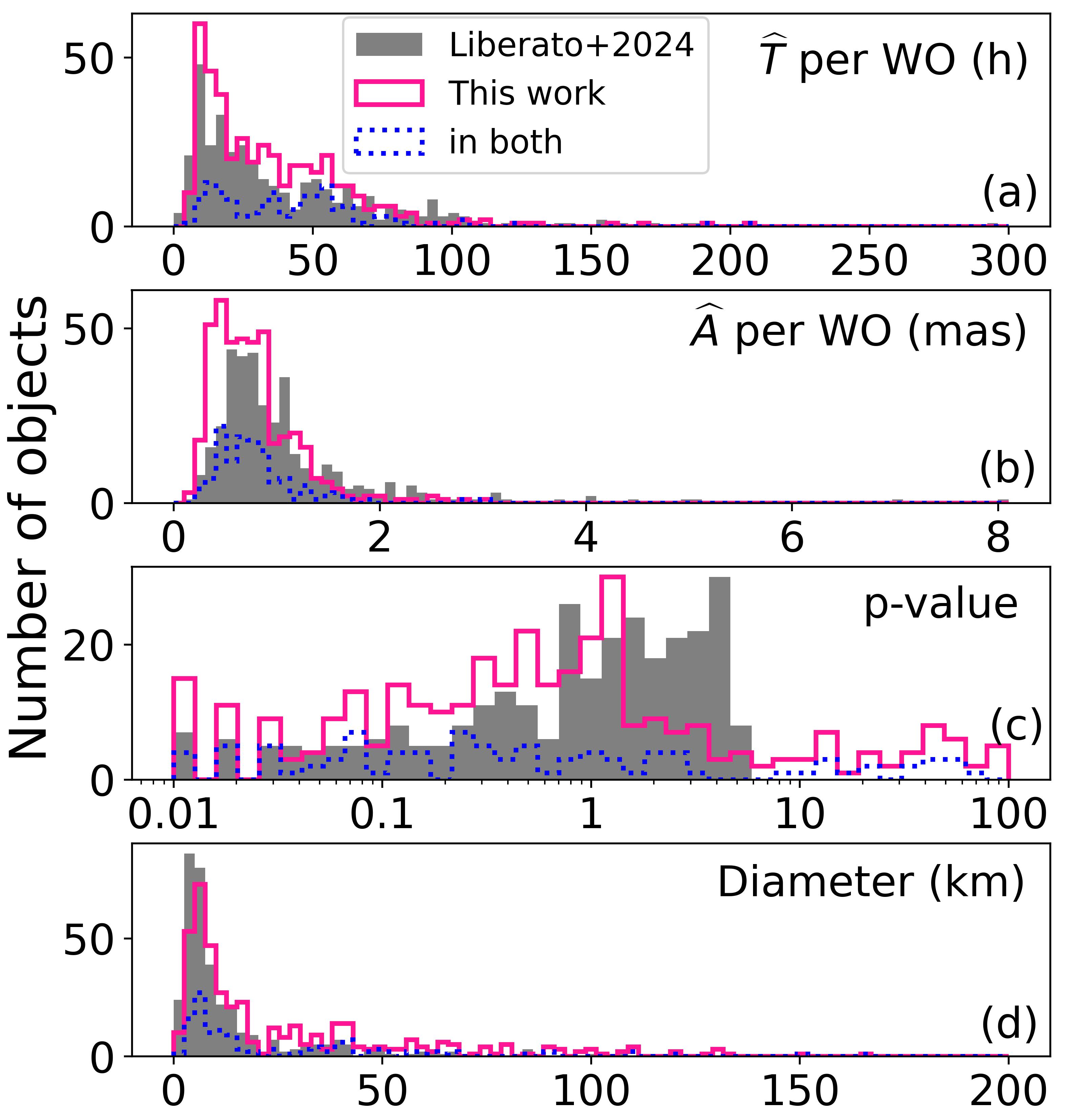}

    \caption{Distribution of period (a), amplitude (b) and p-value in \% (c) estimates per WO, and diameter from the literature (d) for the \gdrthree candidates in L24 (gray bars), for the \gfpr candidates in this work (solid pink line), and the results from the current work for the candidates in common between the two list (dotted blue line).}
    \label{fig:FPR_vs_DR3}
\end{figure}

In Fig.~\ref{fig:FPR_vs_DR3}, we see the comparison of our analysis between the previous and the current binary search. We notice a general qualitative agreement, but small periods and amplitudes are slightly more abundant. We show in (b) that we also have a larger selection of smaller amplitudes, due to the new noise model (Sec.~\ref{sec:errors}). Small wobbles can be due to both smaller sizes and to binaries of similar size ratio. 

We notice in (c) that the current p-values distribution is much more concentrated around small values than in L24, where we adopted a simple threshold at 5\%. This evidence supports the idea that our new noise model allows for clearer detections, along with the FDR control that selects smaller p-values (Sec.~\ref{sec:stat_sel}). Moreover, the majority of the objects present in both lists also present small p-values as seen by the dotted blue distribution in Fig.~\ref{fig:FPR_vs_DR3} (c). Such objects can be considered as those having the strongest wobble signal.

\section{Conclusions}
\label{sec:concl}

In this work, we presented the details of the methodological approach developed and the results of a comprehensive search for astrometric binary asteroids in the \gfpr catalogue. The current method is a substantially improved version of the method presented in L24. The main upgrades include a dedicated noise model for post-fit residuals consistent with the \gaia error model, the identification and detrending of linear systematics in residuals prior to period searches, and a statistically robust selection framework explicitly controlling the false discovery rate (FDR).

The consistency of our FDR threshold with the fraction of objects that are further selected by physical criteria makes us more confident in the reliability of our detections with respect to L24. Still, the 99 objects in common, selected with both methods, are probably the strongest candidates we had in L24. The other asteroids selected in L24 and not appearing here should be considered as weaker, with respect to our current, updated sample.

Finally, we obtain from \gfpr a list of 410 WOs accounting for a total of 343 binary asteroid candidates, representing $\sim$24\% of the 1448 objects initially statistically selected, which is consistent with the expected real detections at FDR threshold of 75\%. We are thus confident that the detection of a periodic signal in the residuals is very solid for a significant fraction of these asteroids. 
In some cases, especially for the largest objects in our sample, both a single or a binary object could be compatible with this signal (as discussed in Sec.~\ref{subsec:likely_bins}).

The selection of the known binaries illustrates the strengths and the limitations in this method, which strongly depends on the quality of the data, the geometry of the observations and the configuration of the systems. Additionally, there are 45 objects with clear trends detected in more than one WO that can correspond to longer periods than those searched for by our approach. 

A significant fraction of our candidates could also be synchronous and fall preferentially, since the wobble period is equivalent to the rotation period of the potential primary. They tend to be more frequent above $\sim$10~km in size, extending into the ``binary desert'' found by other techniques. We now have stronger evidence that \gaia is able to extend the binary detection to an unexplored domain.

Part of the candidate binaries revealed by \gaia overlap with binaries discovered by the mutual events observed in light curves, so they are good candidates for photometry. Systematic surveys in the next years should provide an unprecedentedly large amount of high-quality observations, such as in the case of LSST \citep{kurlander2025predictions,greenstreet2026lightcurves}. Also, all our candidates are excellent targets for stellar occultations, with coordinated campaigns\footnote{Stellar occultation predictions for our candidates are available in https://gaiamoons.imcce.fr/}\citep{lallemand2025sf2a}.%,lallemand2026}.

Besides those techniques, there are not many studies that can be compared to our results. We mention here in particular \citet{ou2022}, where the Point Spread Function of Pan-STARRS1 asteroid observations are analysed. Interestingly, from their 2930 suspected binaries, there are 677 objects with at least one WO in our \gfpr sample. We select 25 of them ($\sim$3.7\%) as binary candidates, including the recently confirmed (720) Bohlinia. 

In future, we intend to apply our revised approach to the Data Release 4, covering double number of asteroids. The revised astrometry in \gdrfour should allow us to consolidate and probably expand our candidate list. 

%%%%%%%%%%%%%%%%%%%%%%%
\section*{Data availability}

All the lists mentioned
can be downloaded from the Zenodo repository: https://doi.org/10.5281/zenodo.18675577

 \begin{acknowledgements}
This work presents results from the European Space Agency (ESA) space mission \gaia. \gaia data are being processed by the \gaia Data Processing and Analysis Consortium (DPAC). Funding for the DPAC is provided by national institutions, in particular, the institutions participating in the \gaia Multilateral Agreement (MLA). The \gaia mission website is https://www.cosmos.esa.int/\gaia. The \gaia archive website is https://archives.esac.esa.int/\gaia.

This work was supported by the project GaiaMoons of the Agence Nationale de Recherche (France), grant ANR-22-CE49-0002. It was financed in part by the French Programme National de Planetologie, and by the BQR program of Observatoire de la C\^ote d'Azur. The authors acknowledge the support by the French National program SUN, project TENET. We made use of the software products: SsODNet VO service of LTE, Observatoire de Paris \citep{berthier2022ssodnet}; Astropy, a community-developed core Python package for Astronomy \citep{1astropy2018, 2astropy2022}; Matplotlib \citep{matplotlib_Hunter:2007}; Multiprocess package \citep{mckerns2010multiprocess,mckerns2012multiprocess}. The authors also thank the valuable contributions of Federica Spoto, Dagmara Oszkiewicz and Aurelie Duchamps.
\end{acknowledgements}

%This research has made use of the Asteroid Families Portal maintained at the Department of Astronomy/University of Belgrade.

%% For this sample we use BibTeX plus aasjournals.bst to generate the
%% the bibliography. The sample631.bib file was populated from ADS. To
%% get the citations to show in the compiled file do the following:
%%
%% pdflatex sample631.tex
%% bibtext sample631
%% pdflatex sample631.tex
%% pdflatex sample631.tex
\bibliographystyle{aa}
\bibliography{references}

@ARTICLE{zechmeister2009,
       author = {{Zechmeister}, M. and {K{\"u}rster}, M.},
        title = "{The generalised Lomb-Scargle periodogram. A new formalism for the floating-mean and Keplerian periodograms}",
      journal = {\aap},
         year = 2009,
        month = mar,
       volume = {496},
       number = {2},
        pages = {577-584},
          doi = {10.1051/0004-6361:200811296},
archivePrefix = {arXiv},
       eprint = {0901.2573},
 primaryClass = {astro-ph.IM},
       adsurl = {https://ui.adsabs.harvard.edu/abs/2009A&A...496..577Z}
}

@article{lomb1976,
  title={Least-squares frequency analysis of unequally spaced data},
  author={Lomb, Nicholas R},
  journal={Astrophysics and space science},
  volume={39},
  number={2},
  pages={447--462},
  year={1976},
  publisher={Springer}
}

@article{scargle1982,
  title={Studies in astronomical time series analysis. II-Statistical aspects of spectral analysis of unevenly spaced data},
  author={Scargle, Jeffrey D},
  journal={ApJ},
  volume={263},
  pages={835--853},
  year={1982}
}

@article{efron1979,
 ISSN = {00905364, 21688966},
 URL = {http://www.jstor.org/stable/2958830},
 author = {B. Efron},
 journal = {The Annals of Statistics},
 number = {1},
 pages = {1--26},
 publisher = {Institute of Mathematical Statistics},
 title = {Bootstrap Methods: Another Look at the Jackknife},
 urldate = {2026-02-11},
 volume = {7},
 year = {1979}
}

@inbook{efron_2010, place={Cambridge}, series={Institute of Mathematical Statistics Monographs}, title={Significance Testing Algorithms}, DOI={10.1017/CBO9780511761362.004}, booktitle={Large-Scale Inference: Empirical Bayes Methods for Estimation, Testing, and Prediction}, publisher={Cambridge University Press}, author={Efron, Bradley}, year={2010}, pages={30–45}, collection={Institute of Mathematical Statistics Monographs}}

@article{pozzi2012,
  title={Exponential smoothing weighted correlations},
  author={Pozzi, Francesco and Di Matteo, Tiziana and Aste, Tomaso},
  journal={The European Physical Journal B},
  volume={85},
  number={6},
  pages={175},
  year={2012},
  publisher={Springer}
}

@article{bertini2012,
  title={Search for satellites near (21) Lutetia using OSIRIS/Rosetta images},
  author={Bertini, Ivano and Sabolo, Walter and Gutierrez, Pedro J and Marzari, Francesco and Snodgrass, Colin and Tubiana, Cecilia and Moissl, Richard and Pajola, Maurizio and Lowry, Stephen C and Barbieri, Cesare and others},
  journal={Planetary and Space Science},
  volume={66},
  number={1},
  pages={64--70},
  year={2012},
  publisher={Elsevier}
}

@article{dell2012,
  title={Observations of asteroids on the Gaia astrometric focal plane},
  author={Dell'Oro, Aldo and Cellino, A},
  journal={Planetary and Space Science},
  volume={73},
  number={1},
  pages={10--14},
  year={2012},
  publisher={Elsevier}
}

@article{kaasalainen2004,
  title={Photocentre offset in ultraprecise astrometry: Implications for barycentre determination and asteroid modelling},
  author={Kaasalainen, M and Tanga, P},
  journal={A\&A},
  volume={416},
  number={1},
  pages={367--373},
  year={2004},
  publisher={EDP Sciences}
}

@book{david2004,
  title={Order statistics},
  author={David, Herbert A and Nagaraja, Haikady N},
  year={2004},
  publisher={John Wiley \& Sons}
}

@article{gorshanov2025,
  title={Testing the binarity of asteroid (720) Bohlinia using lightcurve analysis},
  author={Gorshanov, Denis L and Sokova, Iraida A and Petrova, Svetlana N and Naumov, Konstantin N and Aliev, Amir Kh},
  journal={Planetary and Space Science},
  pages={106217},
  year={2025},
  publisher={Elsevier}
}

@book{milani2010,
  title={Theory of orbit determination},
  author={Milani, Andrea and Gronchi, Giovanni},
  year={2010},
  publisher={Cambridge University Press}
}

@ARTICLE{DAMIT2010,
       author = {{Durech}, J. and {Sidorin}, V. and {Kaasalainen}, M.},
        title = "{DAMIT: a database of asteroid models}",
      journal = {\aap},
         year = 2010,
        month = apr,
       volume = {513},
          eid = {A46},
        pages = {A46},
          doi = {10.1051/0004-6361/200912693},
       adsurl = {https://ui.adsabs.harvard.edu/abs/2010A&A...513A..46D}
}

@article{vanderplas2018,
doi = {10.3847/1538-4365/aab766},
url = {https://dx.doi.org/10.3847/1538-4365/aab766},
year = {2018},
month = {may},
publisher = {The American Astronomical Society},
volume = {236},
number = {1},
pages = {16},
author = {Jacob T. VanderPlas},
title = {Understanding the Lomb–Scargle Periodogram},
journal = {ApJ Supplement Series}
}

@article{carry2010physical,
  title={Physical properties of the ESA Rosetta target asteroid (21) Lutetia-II. Shape and flyby geometry},
  author={Carry, Beno{\i}t and Kaasalainen, Mikko and Leyrat, C{\'e}dric and Merline, Willam J and Drummond, Jack D and Conrad, Al and Weaver, Harold A and Tamblyn, Peter M and Chapman, Clark R and Dumas, Christophe and others},
  journal={A\&A},
  volume={523},
  pages={A94},
  year={2010},
  publisher={EDP Sciences}
}

@article{sierks2011images,
  title={Images of asteroid 21 Lutetia: a remnant planetesimal from the early solar system},
  author={Sierks, Holger and Lamy, Philippe and Barbieri, Cesare and Koschny, Detlef and Rickman, Hans and Rodrigo, Rafael and A’Hearn, Michael F and Angrilli, F and Barucci, M Antonella and Bertaux, J-L and others},
  journal={science},
  volume={334},
  number={6055},
  pages={487--490},
  year={2011},
  publisher={American Association for the Advancement of Science}
}

@article{carry2012density,
  title={Density of asteroids},
  author={Carry, Benoit},
  journal={Planetary and Space Science},
  volume={73},
  number={1},
  pages={98--118},
  year={2012},
  publisher={Elsevier}
}

@Article{matplotlib_Hunter:2007,
  Author    = {Hunter, J. D.},
  Title     = {Matplotlib: A 2D graphics environment},
  Journal   = {Computing in Science \& Engineering},
  Volume    = {9},
  Number    = {3},
  Pages     = {90--95},
  publisher = {IEEE COMPUTER SOC},
  doi       = {10.1109/MCSE.2007.55},
  year      = 2007
}

@article{mckerns2012multiprocess,
  title={Building a framework for predictive science},
  author={McKerns, Michael M and Strand, Leif and Sullivan, Tim and Fang, Alta and Aivazis, Michael AG},
  journal={arXiv preprint arXiv:1202.1056},
  year={2012}
}

@article{mckerns2010multiprocess,
  title={pathos: a framework for heterogeneous computing, 2010},
  author={McKerns, Michael and Aivazis, Michael},
  journal={URL http://trac. mystic. cacr. caltech. edu/project/pathos},
  year={2010}
}

@article{pravec2007binary,
  title={Binary asteroid population: 1. Angular momentum content},
  author={Pravec, Petr and Harris, Alan W},
  journal={Icarus},
  volume={190},
  number={1},
  pages={250--259},
  year={2007},
  publisher={Elsevier}
}

@article{1astropy2018,
       author = {{Astropy Collaboration} and {Price-Whelan}, A.~M. and P. and {Droettboom}, M. and {Bray}, E. and {Aldcroft} and {Sip{\H{o}}cz}},
        title = "{The Astropy Project: Building an Open-science Project and Status of the v2.0 Core Package}",
      journal = {\aj},
         year = 2018,
        month = sep,
       volume = {156},
       number = {3},
          eid = {123},
        pages = {123},
          doi = {10.3847/1538-3881/aabc4f},
archivePrefix = {arXiv},
       eprint = {1801.02634},
 primaryClass = {astro-ph.IM},
       adsurl = {https://ui.adsabs.harvard.edu/abs/2018AJ....156..123A}
}

@ARTICLE{2astropy2022,
       author = {{Astropy Collaboration} and {Price-Whelan}, Adrian M. and {Lim}, Pey Lian and {Earl}, Nicholas and {Starkman}, Nathaniel and {Bradley}, Larry and {Shupe}},
        title = "{The Astropy Project: Sustaining and Growing a Community-oriented Open-source Project and the Latest Major Release (v5.0) of the Core Package}",
      journal = {apj},
         year = 2022,
        month = aug,
       volume = {935},
       number = {2},
          eid = {167},
        pages = {167},
          doi = {10.3847/1538-4357/ac7c74},
archivePrefix = {arXiv},
       eprint = {2206.14220},
 primaryClass = {astro-ph.IM},
       adsurl = {https://ui.adsabs.harvard.edu/abs/2022ApJ...935..167A}
}

@ARTICLE{merline2003esclangona,
       author = {{Merline}, W.~J. and {Close}, L.~M. and {Tamblyn}, P.~M. and {Menard}, F. and {Chapman}, C.~R. and {Dumas}, C. and {Duvert}, G. and {Owen}, W.~M. and {Slater}, D.~C. and {Sterzik}, M.~F.},
        title = "{S/2003 (1509) 1}",
      journal = {\iaucirc},
         year = 2003,
        month = feb,
       volume = {8075},
       adsurl = {https://ui.adsabs.harvard.edu/abs/2003IAUC.8075....2M}
}

@ARTICLE{brown2006,
       author = {M. E. Brown and T.-A. Suer},
        title = "{(55637) 2002 UX\_25}",
      journal = {\iaucirc},
         year = 2007,
        month = feb,
       volume = {8812}
       }

@article{brown2013,
  title={The density of mid-sized Kuiper belt object 2002 UX25 and the formation of the dwarf planets},
  author={Brown, Michael E},
  journal={ApJ Letters},
  volume={778},
  number={2},
  pages={L34},
  year={2013},
  publisher={IOP Publishing}
}

@article{benjamini1995,
  title={Controlling the false discovery rate: a practical and powerful approach to multiple testing},
  author={Benjamini, Yoav and Hochberg, Yosef},
  journal={Journal of the Royal statistical society: series B (Methodological)},
  volume={57},
  number={1},
  pages={289--300},
  year={1995},
  publisher={Wiley Online Library}
}

@ARTICLE{Benishek2023,
       author = {{Benishek}, V. and {Pravec}, P. and {Pilcher}, F.},
        title = "{(2871) Schober}",
      journal = {Central Bureau Electronic Telegrams (CBET) No. 5215},
         year = 2023,
        month = feb,
       volume = {5215}
}

@article{Benishek2025,
  author       = {V. Benishek and P. Pravec and P. Kusnirak and P. Fatka and 
                  R. Durkee and F. Pilcher and K. Ergashev and O. Burkhonov and 
                  D. Augustin and R. Behrend and R. Goncalves},
  title        = {(3220) Murayama},
        journal = {Central Bureau Electronic Telegrams (CBET) No. 5507},
         year = {2025},
        month = {February},
       adsurl = {http://www.cbat.eps.harvard.edu/iau/cbet/005500/CBET005507.txt},
}

@inproceedings{monteiro2024,
  title={Detection and physical characterization of binary asteroids from the IMPACTON project},
  author={Monteiro, Filipe and Oey, Julian and Pereira, Weslley and Evangelista-Santana, Mar{\c{c}}al and Rond{\'o}n, Eduardo and Arcoverde, Plicida and Michimani, Jonatan and Rodrigues, Teresinha and Lazzaro, Daniela},
  booktitle={European Planetary Science Congress},
  pages={EPSC2024--156},
  year={2024}
}

@inproceedings{lallemand2025sf2a,
  title={First GAIAMOONS occultation results with features suggestive of binarity},
  author={Lallemand, R and Desmars, J and Sicardy, B and Liu, Z and Liberato, L and Tanga, P},
  booktitle={SF2A-2025: Proceedings of the Annual meeting of the French Society of Astronomy and Astrophysics. Eds.: A. Siebert},
  pages={177--180},
  year={2025}
}

@ARTICLE{Benishek2024a,
  author       = "V. Benishek and P. Pravec and A. Marchini and R. Papini and F. Pilcher and N. Ruocco and R. Durkee",
  title        = "(1879) Broederstroom",
  journal = "{Central Bureau Electronic Telegrams (CBET) No. 5373}",
  year         = "2024",
  month        = "March",
  url          = "http://www.cbat.eps.harvard.edu/iau/cbet/005300/CBET005373.txt"
}

@article{Pray2015,
  author = {D. Pray and P. Pravec and K. Hornoch and J. Vrastil and H. Kucakova and V. Benishek and R. Roy and R. Behrend and D. Romeuf and B. Warner and A. Scholz and G. Hodosan and J. Pollock and R. Montaigut and A. Leroy and D. Reichart and J. Haislip},
  year = {2015},
  journal={Central Bureau Electronic Telegrams (CBET) No. 4157}
}

@article{Sato2013,
  author       = {Isao Sato},
  title        = {日本人による多重小惑星の発見 = Detections of Multiple Asteroids from Japan},
  journal      = {Spaceguard Research},
  volume       = {5},
  pages        = {23--25},
  year         = {2013},
  note         = {\textit{Spaceguard Research}, Vol.\ 5},
  url          = {https://www.spaceguard.or.jp/RSGC/results/sgr5pdfs/23_Sato.pdf}
}

@article{Garrison_2024,
  doi = {10.3847/2515-5172/ad82cd},
  url = {https://dx.doi.org/10.3847/2515-5172/ad82cd},
  year = {2024},
  month = {oct},
  publisher = {The American Astronomical Society},
  volume = {8},
  number = {10},
  pages = {250},
  author = {Lehman H. Garrison and Dan Foreman-Mackey and Yu-hsuan Shih and Alex Barnett},
  title = {nifty-ls: Fast and Accurate Lomb–Scargle Periodograms Using a Non-uniform FFT},
  journal = {Research Notes of the AAS}
}

@article{kurlander2025predictions,
  title={Predictions of the LSST solar system yield: Near-Earth objects, main belt asteroids, Jupiter Trojans, and trans-Neptunian objects},
  author={Kurlander, Jacob A and Bernardinelli, Pedro H and Schwamb, Megan E and Juri{\'c}, Mario and Murtagh, Joseph and Chandler, Colin Orion and Merritt, Stephanie R and Nesvorn{\`y}, David and Vokrouhlick{\`y}, David and Jones, R Lynne and others},
  journal={AJ},
  volume={170},
  number={2},
  pages={99},
  year={2025},
  publisher={The American Astronomical Society}
}

@article{greenstreet2026lightcurves,
  title={Lightcurves, Rotation Periods, and Colors for Vera C. Rubin Observatory’s First Asteroid Discoveries},
  author={Greenstreet, Sarah and Li, Zhuofu Chester and Vavilov, Dmitrii E and Singh, Devanshi and Juri{\'c}, Mario and Ivezi{\'c}, {\v{Z}}eljko and Eggl, Siegfried and Koumjian, Alec and Moeyens, Joachim and Carruba, Valerio and others},
  journal={ApJ Letters},
  volume={996},
  number={2},
  pages={L33},
  year={2026},
  publisher={IOP Publishing}
}

@ARTICLE{berthier2022ssodnet,
       author = {{Berthier}, J. and {Carry}, B. and {Mahlke}, M. and {Normand}, J.},
        title = "{SsODNet: Solar system Open Database Network}",
      journal = {\aap},
         year = 2023,
        month = mar,
       volume = {671},
          eid = {A151},
        pages = {A151},
          doi = {10.1051/0004-6361/202244878},
archivePrefix = {arXiv},
       eprint = {2209.10697},
 primaryClass = {astro-ph.EP},
       adsurl = {https://ui.adsabs.harvard.edu/abs/2023A&A...671A.151B}
}

@article{JArchive,
  author    = "Johnston, W. R.",
  year      = "2025",
  journal   = "Asteroids with Satellites URL: www.johnstonsarchive.net/astro/asteroidmoons.html",
  note      = "Accessed in September, 2025"
}

@incollection{hestroffer2010gaia,
  title={The Gaia mission and the asteroids},
  author={Hestroffer, Daniel and Cellino, Alberto and Tanga, Paolo and others},
  booktitle={Dynamics of Small Solar System Bodies and Exoplanets},
  pages={251--340},
  year={2010},
  publisher={Springer}
}

@article{pravec2012small,
  title={Small binary asteroids and prospects for their observations with Gaia},
  author={Pravec, P and Scheirich, P},
  journal={Planetary and Space Science},
  volume={73},
  number={1},
  pages={56--61},
  year={2012},
  publisher={Elsevier}
}

@article{margot2015,
  title={Asteroid systems: binaries, triples, and pairs},
  author={Margot, Jean-Luc and Pravec, Petr and Taylor, Patrick and Carry, Beno{\i}t and Jacobson, Seth},
  journal={Asteroids IV},
  volume={355},
  pages={373},
  year={2015},
  publisher={University of Arizona Press Tucson}
}

@article{durda2004,
  title={The formation of asteroid satellites in large impacts: Results from numerical simulations},
  author={Durda, Daniel D and Bottke Jr, William F and Enke, Brian L and Merline, William J and Asphaug, Erik and Richardson, Derek C and Leinhardt, Zo{\"e} M},
  journal={Icarus},
  volume={167},
  number={2},
  pages={382--396},
  year={2004},
  publisher={Elsevier}
}

@article{cellino2024asteroid,
  title={Asteroid spin and shape properties from Gaia DR3 photometry},
  author={Cellino, A and Tanga, P and Muinonen, K and Mignard, F},
  journal={A\&A},
  volume={687},
  pages={A277},
  year={2024},
  publisher={EDP sciences}
}

@article{liberato2026zenodo,
  title        = {Follow the wobble: Statistical methods to detect astrometric binary asteroids in Gaia FPR},
  author       = {Liberato, Luana and Tanga, Paolo and Mary, David and Lallemand, Raphael and Liu, Ziyu and Desmars, Josselin and Hestroffer, Daniel and Carry, Benoit and Minker, Kate and Siakas, Alexandros},

  year = 2026,
  doi          = {10.5281/zenodo.18675577},
  volume          = {https://doi.org/10.5281/zenodo.18675577},
  journal    = {Zenodo repository}

}

@article{heard2018,
  title={Choosing between methods of combining-values},
  author={Heard, Nicholas A and Rubin-Delanchy, Patrick},
  journal={Biometrika},
  volume={105},
  number={1},
  pages={239--246},
  year={2018},
  publisher={Oxford University Press}
}

@article{vovk2020,
  title={Combining p-values via averaging},
  author={Vovk, Vladimir and Wang, Ruodu},
  journal={Biometrika},
  volume={107},
  number={4},
  pages={791--808},
  year={2020},
  publisher={Oxford University Press}
}

@article{drummond2021orbit,
  title={The orbit of asteroid (317) Roxane’s satellite Olympias from Gemini, Keck, VLT and the SOR, and (22) Kalliope’s Linus from the SOR},
  author={Drummond, Jack D and Merline, WJ and Carry, B and Conrad, A and Tamblyn, P and Enke, B and Christou, J and Dumas, C and Chapman, CR and Durda, DD and others},
  journal={Icarus},
  volume={358},
  pages={114275},
  year={2021},
  publisher={Elsevier}
}

@misc{tippett1931,
  title={The methods of Statistics, Williams and Norgate, London; 1952},
  author={Tippett, LHC},
  year={1931},
  publisher={Wiley, New York}
}

@article{fisher1948,
  title={Combining independent tests of significance},
  author={Fisher, Ronald A},
  journal={American Statistician},
  volume={2},
  number={5},
  pages={30},
  year={1948}
}

@article{scheeres2015asteroid,
  title={Asteroid interiors and morphology},
  author={Scheeres, DJ and Britt, D and Carry, B and Holsapple, KA},
  journal={Asteroids iv},
  volume={745766},
  pages={745--766},
  year={2015},
  publisher={University of Arizona Press Tucson, Arizona}
}

@Article{durech2003,
  author  = {{\v D}urech, J. and Kaasalainen, M.},
  title   = {{Photometric signatures of highly nonconvex and binary asteroids}},
  journal = {\aap},
  year    = {2003},
  volume  = {404},
  pages   = {709--714},
  month   = jun,
  adsurl  = {n.a.},
  doi     = {10.1051/0004-6361:20030505},
}

@article{tanga2023,
  title={Gaia Data Release 3-The Solar System survey},
  author={Tanga, P and Pauwels, T and Mignard, F and Muinonen, K and Cellino, A and David, P and Hestroffer, D and Spoto, F and Berthier, J and Guiraud, J and others},
  journal={A\&A},
  volume={674},
  pages={A12},
  year={2023},
  publisher={EDP sciences}
}

@article{demeo2015compositional,
  title={The compositional structure of the asteroid belt},
  author={DeMeo, FE and Alexander, CMO and Walsh, KJ and Chapman, CR and Binzel, RP and others},
  journal={Asteroids iv},
  volume={1},
  number={3},
  pages={13--41},
  year={2015},
  publisher={University of Arizona Press Tucson, AZ}
}

@article{izidoro2015terrestrial,
  title={Terrestrial planet formation constrained by Mars and the structure of the asteroid belt},
  author={Izidoro, Andr{\'e} and Raymond, Sean N and Morbidelli, Alessandro and Winter, Othon C},
  journal={MNRAS},
  volume={453},
  number={4},
  pages={3619--3634},
  year={2015},
  publisher={Oxford University Press}
}

@article{morbidelli2015dynamical,
  title={The dynamical evolution of the asteroid belt},
  author={Morbidelli, Alessandro and Walsh, Kevin J and O'Brien, David P and Minton, David A and Bottke, William F},
  journal={arXiv preprint arXiv:1501.06204},
  year={2015}
}

@article{ou2022,
  title={Searching for Binary Asteroids in Pan-STARRS1 Archival Images},
  author={Ou, James and Baranec, Christoph and Bus, Schelte J},
  journal={The Planetary Science Journal},
  volume={3},
  number={7},
  pages={169},
  year={2022},
  publisher={IOP Publishing}
}

@article{gaiafpr,
  title={Gaia Focused Product Release: Asteroid orbital solution-Properties and assessment},
  author={David, Patrice and Mignard, Fran{\c{c}}ois and Hestroffer, Daniel and Tanga, Paolo and Spoto, Federica and Berthier, J{\'e}r{\^o}me and Pauwels, Thierry and Roux, Wilhem and Barbier, A and Cellino, Alberto and others},
  journal={A\&A},
  volume={680},
  pages={A37},
  year={2023},
  publisher={EDP sciences}
}

@article{spoto2018,
  title={Gaia Data Release 2-Observations of solar system objects},
  author={Spoto, Federica and Tanga, P and Mignard, Fran{\c{c}}ois and Berthier, Jean and Carry, B and Cellino, A and Dell’Oro, A and Hestroffer, D and Muinonen, K and Pauwels, T and others},
  journal={A\&A},
  volume={616},
  pages={A13},
  year={2018},
  publisher={EDP sciences}
}

@article{gaiacollaborationDR2_2018,
  title = {Gaia {{Data Release}} 2. {{Observations}} of Solar System Objects},
  author = {{Gaia Collaboration} and Spoto, F. and Tanga, P. and Mignard, F. and Berthier, J. and Carry, B. and Cellino, A. and Dell'Oro, A. and Hestroffer, D. and Muinonen, K. and Pauwels, T. and Petit, J.-M. and David, P. and De Angeli, F. and Delbo, M. and Fr{\'e}, B. and Galluccio, L. and Granvik, M. and Guiraud, J. and Hern{\'a}, J. and Ord{\'e}, C. and Portell, J. and Poujoulet, E. and Thuillot, W. and Walmsley, G. and Brown, A. G. A. and Vallenari, A. and Prusti, T. and {de Bruijne}, J. H. J. and Babusiaux, C. and {Bailer-Jones}, C. A. L. and Biermann, M. and Evans, D. W. and Eyer, L. and Jansen, F. and Jordi, C. and Klioner, S. A. and Lammers, U. and Lindegren, L. and Luri, X. and Panem, C. and Pourbaix, D. and Randich, S. and Sartoretti, P. and Siddiqui, H. I. and Soubiran, C. and {van Leeuwen}, F. and Walton, N. A. and Arenou, F. and Bastian, U. and Cropper, M. and Drimmel, R. and Katz, D. and Lattanzi, M. G. and Bakker, J. and Cacciari, C. and Casta{\~n}, J. and Chaoul, L. and Cheek, N. and Fabricius, C. and Guerra, R. and Holl, B. and Masana, E. and Messineo, R. and Mowlavi, N. and Nienartowicz, K. and Panuzzo, P. and Riello, M. and Seabroke, G. M. and Th{\'e}, F. and {Gracia-Abril}, G. and Comoretto, G. and {Garcia-Reinaldos}, M. and Teyssier, D. and Altmann, M. and Andrae, R. and Audard, M. and {Bellas-Velidis}, I. and Benson, K. and Blomme, R. and Burgess, P. and Busso, G. and Clementini, G. and Clotet, M. and Creevey, O. and Davidson, M. and De Ridder, J. and Delchambre, L. and Ducourant, C. and Fern{\'a}, J. and Fouesneau, M. and Fr{\'e}, Y. and Garc{\'i}, M. and Gonz{\'a}, J. and Gonz{\'a}, J. J. and Gosset, E. and Guy, L. P. and Halbwachs, J.-L. and Hambly, N. C. and Harrison, D. L. and Hodgkin, S. T. and Hutton, A. and Jasniewicz, G. and {Jean-Antoine-Piccolo}, A. and Jordan, S. and Korn, A. J. and {Krone-Martins}, A. and Lanzafame, A. C. and Lebzelter, T. and {L o{\textasciidieresis}}, W. and Manteiga, M. and Marrese, P. M. and Mart{\'i}, J. M. and Moitinho, A. and Mora, A. and Osinde, J. and Pancino, E. and {Recio-Blanco}, A. and Richards, P. J. and Rimoldini, L. and Robin, A. C. and Sarro, L. M. and Siopis, C. and Smith, M. and Sozzetti, A. and S{\"u}veges, M. and Torra, J. and {van Reeven}, W. and Abbas, U. and Abreu Aramburu, A. and Accart, S. and Aerts, C. and Altavilla, G. and {\'A}lvarez, M. A. and Alvarez, R. and Alves, J. and Anderson, R. I. and Andrei, A. H. and Anglada Varela, E. and Antiche, E. and Antoja, T. and Arcay, B. and Astraatmadja, T. L. and Bach, N. and Baker, S. G. and {Balaguer-N{\'u}}, L. and Balm, P. and Barache, C. and Barata, C. and Barbato, D. and Barblan, F. and Barklem, P. S. and Barrado, D. and Barros, M. and Barstow, M. A. and Bartholom{\'e}, S. and Bassilana, J.-L. and Becciani, U. and Bellazzini, M. and Berihuete, A. and Bertone, S. and Bianchi, L. and Bienaym{\'e}, O. and {Blanco-Cuaresma}, S. and Boch, T. and Boeche, C. and Bombrun, A. and Borrachero, R. and Bossini, D. and Bouquillon, S. and Bourda, G. and Bragaglia, A. and Bramante, L. and Breddels, M. A. and Bressan, A. and Brouillet, N. and Br{\"u}semeister, T. and Brugaletta, E. and Bucciarelli, B. and Burlacu, A. and Busonero, D. and Butkevich, A. G. and Buzzi, R. and Caffau, E. and Cancelliere, R. and Cannizzaro, G. and {Cantat-Gaudin}, T. and Carballo, R. and Carlucci, T. and Carrasco, J. M. and Casamiquela, L. and Castellani, M. and {Castro-Ginard}, A. and Charlot, P. and Chemin, L. and Chiavassa, A. and Cocozza, G. and Costigan, G. and Cowell, S. and Crifo, F. and Crosta, M. and Crowley, C. and Cuypers, J. and Dafonte, C. and Damerdji, Y. and Dapergolas, A. and David, M. and {de Laverny}, P. and De Luise, F. and De March, R. and {de Souza}, R. and {de Torres}, A. and Debosscher, J. and {del Pozo}, E. and Delgado, A. and Delgado, H. E. and Diakite, S. and Diener, C. and Distefano, E. and Dolding, C. and Drazinos, P. and Dur{\'a}, J. and Edvardsson, B. and Enke, H. and Eriksson, K. and Esquej, P. and Eynard Bontemps, G. and Fabre, C. and Fabrizio, M. and Faigler, S. and Falc{\~a}, A. J. and Farr{\`a}, M. and Federici, L. and Fedorets, G. and Fernique, P. and Figueras, F. and Filippi, F. and Findeisen, K. and Fonti, A. and Fraile, E. and Fraser, M. and Gai, M. and Galleti, S. and Garabato, D. and Garc{\'i}, F. and Garofalo, A. and Garralda, N. and Gavel, A. and Gavras, P. and Gerssen, J. and Geyer, R. and Giacobbe, P. and Gilmore, G. and Girona, S. and Giuffrida, G. and Glass, F. and Gomes, M. and Gueguen, A. and Guerrier, A. and Guti{\'e}, R. and Haigron, R. and Hatzidimitriou, D. and Hauser, M. and Haywood, M. and Heiter, U. and Helmi, A. and Heu, J. and Hilger, T. and Hobbs, D. and Hofmann, W. and Holland, G. and Huckle, H. E. and Hypki, A. and Icardi, V. and Jan{\ss}, K. and {Jevardat de Fombelle}, G. and Jonker, P. G. and Juh{\'a}, {\'A}. L. and Julbe, F. and Karampelas, A. and Kewley, A. and Klar, J. and Kochoska, A. and Kohley, R. and Kolenberg, K. and Kontizas, M. and Kontizas, E. and Koposov, S. E. and Kordopatis, G. and {Kostrzewa-Rutkowska}, Z. and Koubsky, P. and Lambert, S. and Lanza, A. F. and Lasne, Y. and Lavigne, J.-B. and Le Fustec, Y. and {Le Poncin-Lafitte}, C. and Lebreton, Y. and Leccia, S. and Leclerc, N. and {Lecoeur-Taibi}, I. and Lenhardt, H. and Leroux, F. and Liao, S. and Licata, E. and Lindstr{\o}, H. E. P. and Lister, T. A. and Livanou, E. and Lobel, A. and L{\'o}, M. and Managau, S. and Mann, R. G. and Mantelet, G. and Marchal, O. and Marchant, J. M. and Marconi, M. and Marinoni, S. and Marschalk{\'o}, G. and Marshall, D. J. and Martino, M. and Marton, G. and Mary, N. and Massari, D. and Matijevi{\v c}, G. and Mazeh, T. and McMillan, P. J. and Messina, S. and Michalik, D. and Millar, N. R. and Molina, D. and Molinaro, R. and Moln{\'a}, L. and Montegriffo, P. and Mor, R. and Morbidelli, R. and Morel, T. and Morris, D. and Mulone, A. F. and Muraveva, T. and Musella, I. and Nelemans, G. and Nicastro, L. and Noval, L. and O'Mullane, W. and Ord{\'o}, D. and Osborne, P. and Pagani, C. and Pagano, I. and Pailler, F. and Palacin, H. and Palaversa, L. and Panahi, A. and Pawlak, M. and Piersimoni, A. M. and Pineau, F.-X. and Plachy, E. and Plum, G. and Poggio, E. and Pr{\v s}, A. and Pulone, L. and Racero, E. and Ragaini, S. and Rambaux, N. and {Ramos-Lerate}, M. and Regibo, S. and Reyl{\'e}, C. and Riclet, F. and Ripepi, V. and Riva, A. and Rivard, A. and Rixon, G. and Roegiers, T. and Roelens, M. and {Romero-G{\'o}}, M. and Rowell, N. and Royer, F. and {Ruiz-Dern}, L. and Sadowski, G. and Sagrist{\`a}, T. and Sahlmann, J. and Salgado, J. and Salguero, E. and Sanna, N. and {Santana-Ros}, T. and Sarasso, M. and Savietto, H. and Schultheis, M. and Sciacca, E. and Segol, M. and Segovia, J. C. and S{\'e}, D. and Shih, I.-C. and Siltala, L. and Silva, A. F. and Smart, R. L. and Smith, K. W. and Solano, E. and Solitro, F. and Sordo, R. and Soria Nieto, S. and Souchay, J. and Spagna, A. and Stampa, U. and Steele, I. A. and Steidelm{\"u}ller, H. and Stephenson, C. A. and Stoev, H. and Suess, F. F. and Surdej, J. and Szabados, L. and {Szegedi-Elek}, E. and Tapiador, D. and Taris, F. and Tauran, G. and Taylor, M. B. and Teixeira, R. and Terrett, D. and Teyssandier, P. and Titarenko, A. and Torra Clotet, F. and Turon, C. and Ulla, A. and Utrilla, E. and Uzzi, S. and Vaillant, M. and Valentini, G. and Valette, V. and {van Elteren}, A. and Van Hemelryck, E. and {van Leeuwen}, M. and Vaschetto, M. and Vecchiato, A. and Veljanoski, J. and Viala, Y. and Vicente, D. and Vogt, S. and {von Essen}, C. and Voss, H. and Votruba, V. and Voutsinas, S. and Weiler, M. and Wertz, O. and Wevers, T. and Wyrzykowski, {\L}. and Yoldas, A. and {\v Z}erjal, M. and Ziaeepour, H. and Zorec, J. and Zschocke, S. and Zucker, S. and Zurbach, C. and Zwitter, T.},
  year = {2018},
  month = aug,
  journal = {Astronomy and Astrophysics},
  volume = {616},
  pages = {A13},
  issn = {0004-6361},
  doi = {10.1051/0004-6361/201832900},
  urldate = {2018-08-20}
}

@article{liberato2024,
  title={Binary asteroid candidates in Gaia DR3 astrometry},
  author={Liberato, Luana and Tanga, Paolo and Mary, David and Minker, Kate and Carry, Beno{\^\i}t and Spoto, Federica and Bartczak, Przemyslaw and Sicardy, Bruno and Oszkiewicz, Dagmara and Desmars, Josselin},
  journal={A\&A},
  volume={688},
  pages={A50},
  year={2024},
  publisher={EDP Sciences}
}

@article{liu2024,
  title={Asteroid (4337) Arecibo: Two ice-rich bodies forming a binary-Based on Gaia astrometric data},
  author={Liu, Ziyu and Hestroffer, Daniel and Desmars, Josselin and David, Pedro},
  journal={A\&A},
  volume={688},
  pages={L23},
  year={2024},
  publisher={EDP Sciences}
}

@article{gault2022new,
  title={A new satellite of 4337 arecibo detected and confirmed by stellar occultation},
  author={Gault, David and Nosworthy, Peter and Nolthenius, Richard and Bender, Kirk and Herald, Dave},
  journal={Minor Planet Bulletin},
  volume={49},
  number={1},
  pages={3--5},
  year={2022}
}

@article{demeo2014solar,
  title={Solar System evolution from compositional mapping of the asteroid belt},
  author={DeMeo, Francesca E and Carry, Beno{\i}t},
  journal={Nature},
  volume={505},
  number={7485},
  pages={629--634},
  year={2014},
  publisher={Nature Publishing Group UK London}
}

@article{kleine2002rapid,
  title={Rapid accretion and early core formation on asteroids and the terrestrial planets from Hf--W chronometry},
  author={Kleine, T and M{\"u}nker, C and Mezger, Klaus and Palme, H},
  journal={Nature},
  volume={418},
  number={6901},
  pages={952--955},
  year={2002},
  publisher={Nature Publishing Group UK London}
}

@article{lindegren2021,
  title={Gaia early data release 3-the astrometric solution},
  author={Lindegren, Lennart and Klioner, SA and Hern{\'a}ndez, J and Bombrun, A and Ramos-Lerate, M and Steidelm{\"u}ller, Hea and Bastian, U and Biermann, M and de Torres, A and Gerlach, E and others},
  journal={A\&A},
  volume={649},
  pages={A2},
  year={2021},
  publisher={EDP Sciences}
}

@article{fuentes2024, 
 title={Asteroid Orbit Determination Using Gaia FPR: Statistical Analysis}, volume={167}, ISSN={0004-6256}, DOI={10.3847/1538-3881/ad4291}, number={6}, journal={AJ, Volume 167, Issue 6, id.290, 9 pp.}, author={Fuentes-Muñoz, Oscar and Farnocchia, Davide and Naidu, Shantanu P. and Park, Ryan S.}, year={2024}, month=june, pages={290}, language={en} }

\appendix
 
\section{Algorithm for confidence interval estimation}
\label{app:algo}
We present here the pseudo-code that summarises the algorithm used in this work to estimate the confidence intervals for the period (${\mathcal{I}}_T$) and amplitude (${\mathcal{I}}_A$) measured from the wobble detected in each WO, as described in more detail in the text of Section \ref{subsec:algo}.

The estimated parameter of the wobble $\widehat{f}$ and $\widehat{A}$, along with the residuals' sample, are the inputs. The variable $M$ indicates the number of Monte-Carlo simulations, while $\mu^{(i)}$ is the systematic offset and $\bn^{(i)}$ is the random noise, added to each of the $M$ time series simulated (steps 2 to 5).

In steps 12 to 14, the notation $X_{(m)}$ denotes the $m$th order statistics of the data distribution $X$ (i.e. the $m$th quantile of $X$), and  $(\lfloor x \rfloor)$ the nearest integer smaller than or equal to $x$. To add robustness to the CI, we compute the distances between two quantiles and the median (step 15) and use the largest one to compute a symmetric interval around $\widehat{A}$.

\IncMargin{1em}
\begin{algorithm}[]
\SetKwInOut{Input}{Inputs}\SetKwInOut{Output}{Output}
\Input{${\boldsymbol{t}}:=[t_1,\cdots,t_K]^\top:$  residuals' epochs\\
${\boldsymbol{{\sigma}}}:=[{\sigma}_1,\cdots,{\sigma}_K]^\top:$ $K$ std of random error on the residuals\\
$\widehat{A},\widehat{f}$ ({and} $\widehat{T}=1/\widehat{f}$) : parameters of the estimated wobble   \\
${\bf{\nu}}:=[\nu_{\min},\cdots,\nu_{\max}]^\top$: vector of frequency search for GLSP \\
$M$: number of MC realisations}
\Output{${\mathcal{I}}_A$ and ${\mathcal{I}}_T$: the $95\%$ CI for ${A}$ and ${T}$}
\BlankLine

\For{$i=1,\cdots,M$}{
Draw random phase: $\varphi^{(i)} \sim \mathcal{U}_{[0,2\pi]}$\;
Draw random offset: $\mu^{(i)} \sim {\mathcal{L}}(0.02,0.19)$ and constant offset vector ${\boldsymbol{\mu}}^{(i)}:=[\mu^{(i)},\cdots,\mu^{(i)}]^\top$\;
Draw random noise vector: ${\bf{n}}^{(i)}\sim \mathcal{N}({\bf{0}},\mathrm{diag}({\boldsymbol{{\sigma}}}))$\;
Generate time series:
${\bf{y}}^{(i)}={\bf{\mu}}^{(i)}+\widehat{A}\sin(2\pi \widehat{f}~ {\bf{t}}+\varphi^{(i)} )+\bn^{(i)}$

Compute GLSP: $\mathcal{P}^{(i)}(\nu;\by^{(i)},{\bsigma}, \bt)$\;
$\widehat{f}^{(i)} \leftarrow \arg\max_\nu \mathcal{P}^{(i)}(\nu)$\;
 $\widehat{T}^{(i)}\leftarrow  1/\widehat{f}^{(i)}$\;
Compute $\widehat{A}^{(i)}$ and  $\widehat{\varphi}^{(i)}$ from WLS fit of a sinusoid with frequency $\widehat{f}^{(i)}$ to $\by^{(i)} $ (Eq. \eqref{eq:beta_wls}).
}
\BlankLine
Sort $\{\widehat{A}^{(i)}\}_{i=1}^M$ and  $\{\widehat{T}^{(i)}\}_{i=1}^M$ in increasing order\;

$q^{\widehat{A}}_{2.5}\leftarrow \widehat{A}^{(i)}_{(\lfloor0.025\times M\rfloor)}$\;
$q^{\widehat{A}}_{50} \leftarrow \widehat{A}^{(i)}_{(\lfloor0.5\times M\rfloor)}$\;
$ q^{\widehat{A}}_{97.5} \leftarrow \widehat{A}^{(i)}_{(\lfloor0.975\times M\rfloor)}$\;
 $q^{\widehat{A}\star}:=\max \{q^{\widehat{A}}_{97.5}-q^{\widehat{A}}_{50},q^{\widehat{A}}_{50}-q^{\widehat{A}}_{2.5} \}$\;
Compute the $95\%$ CI for $\widehat{A}$: ${\mathcal{I}}_{A} \leftarrow [\widehat{A}-q^{\widehat{A}\star},\widehat{A}+q^{\widehat{A}\star}]$\; 

Repeat steps 12--16 applied to $\{\widehat{T}^{(i)}\}_{i=1}^M$ to produce ${\mathcal{I}}_{{T}}$.

\caption{Estimation of confidence intervals.}
\label{algo:ci_fpr}
\end{algorithm}

\section{Pearson's correlation coefficient}
\label{app:pearson}
One of the most used methods to quantify the strength and direction of a linear relationship between two variables is Pearson's correlation coefficient. If different realisations of the variables have different reliabilities or uncertainties, a weighted version is more useful since it allows more precise measurements to contribute more strongly, while still preventing noisy points from dominating the correlation.

The conventional weighted Pearson's correlation coefficient \citep[see Sec. 3.1 of ][]{pozzi2012} can be obtained using the following equation:

\begin{equation}
c({\bf t},{\bf y}) \;=\;\frac{  \displaystyle \sum_{k=1}^K w_k \, (t_k - m_t)(y_k - m_y)}
{ \displaystyle \sqrt{\left( \sum_{k=1}^K w_k (t_k - m_t)^2 \right)
              \left( \sum_{k=1}^K w_k (y_k - m_y)^2 \right)}}
\end{equation}

with:

\begin{equation}
w_k  := \frac{1}{\sigma_{k}^2}, 
\qquad 
m_t := \frac{\sum_{k=1}^K w_k t_k}{\sum_{k=1}^K w_k},
\qquad
m_y := \frac{\sum_{k=1}^K w_k y_k}{\sum_{k=1}^K w_k},
\end{equation}

where the weights $w_k$ are the inverse noise variances  (see Sec.~\ref{sec:errors}), ${\bf{t}}=[t_1,\cdots,t_K]^\top$ are the transit averaged epochs in one WO and ${\bf{y}}=[y_1,\cdots,y_K]^\top$ the corresponding residuals.  
%****************************************************************

\section{GLSP with general linear model}
\label{app:gglsp}

\begin{figure}[ht]
    \centering
        \includegraphics[width=0.98\linewidth]{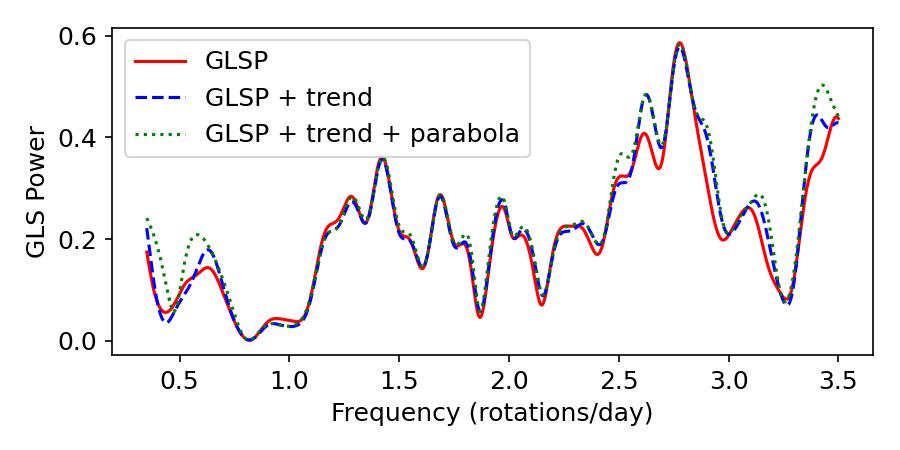}\\
        \includegraphics[width=1.01\linewidth]{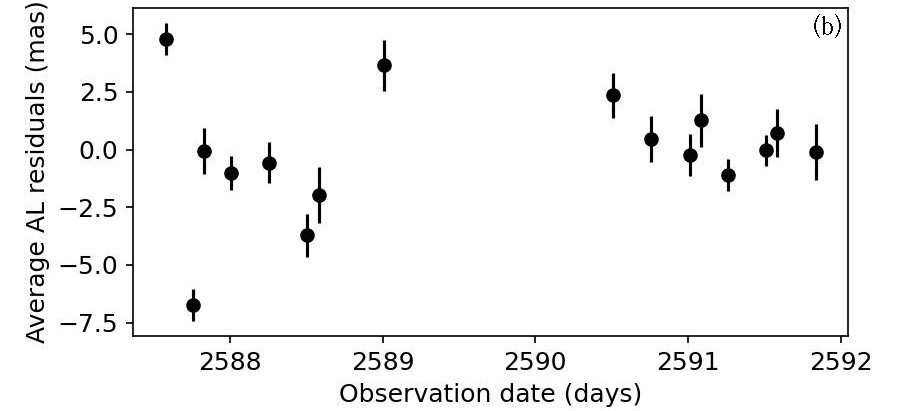}\\
        \includegraphics[width=0.98\linewidth]{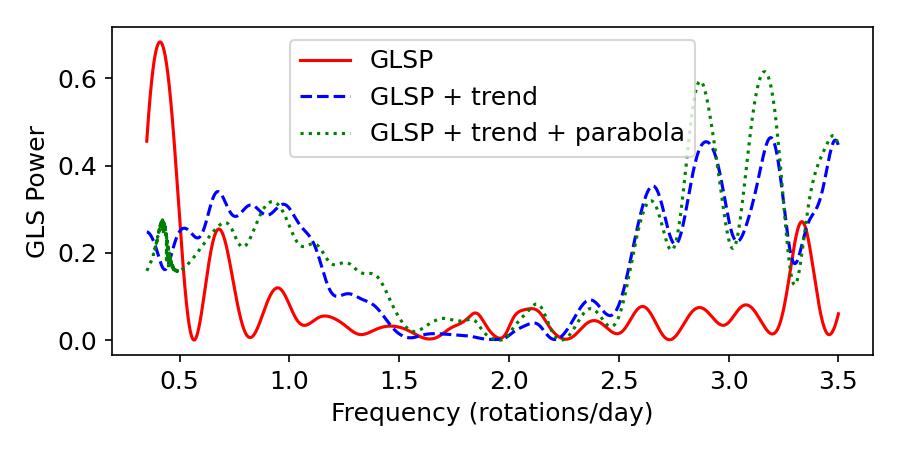}\\
        \includegraphics[width=0.98\linewidth]{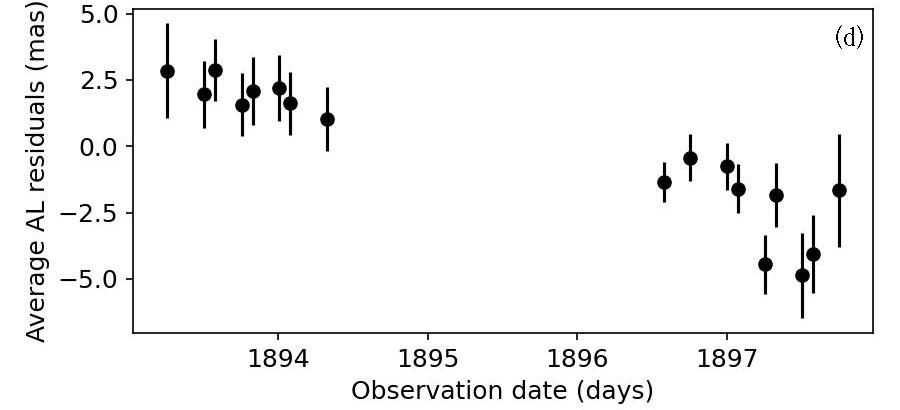}
   
    \caption{Comparison between classical GLSP (red solid) and GLSP including constant +  linear dashed blue) and constant + linear + quadratic (dotted green) for two different samples of residuals explored. Plots (a) and (c) show the periodograms for samples (b) and (d), respectively.}
    \label{fig:sameGLSP}
\end{figure}

Consider a general linear model for the time series of the residuals per transit :
\begin{equation}
\by = \bM \bbet + \boldsymbol{\epsilon}
\label{modelGLSP}
\end{equation}
with $\bM$ a model matrix, $\bbet$ the coefficients associated with each column of $\bM$,  and $\boldsymbol{\epsilon} \sim {\cal{N}}({\bf{0}}, \boldsymbol{\Sigma})$ the covariance matrix of the noise. 
For instance, for a ``sinusoid + constant model'', the matrix is:
\begin{equation}
\label{Mnu}
    \bM(\nu):=[{\bf{1}}, \; \bcc(\nu), \; \bss(\nu)  ] 
\end{equation}
    with 
 \begin{eqnarray}
    {\bf{1}}& :=&[ 1, \cdots, 1 ]^\top,\label{B3}\\ 
    \bcc(\nu) &:=& [ \cos(2\pi \nu t_1),\cdots , \cos(2\pi \nu t_K)  ]^\top, \\ 
    \bss(\nu) &:=& [ \sin(2\pi \nu  t_1),\cdots, \sin(2\pi \nu t_K)  ]^\top.
\end{eqnarray}
In case the noise is assumed uncorrelated (as this is the case for the $K$ \gaia residuals within a WO), but with a different variance $\sigma^2_i$ on each sample, then  the matrix $\boldsymbol{\Sigma}  $ is diagonal with diagonal $\textrm{diag}; \boldsymbol{\Sigma}  =[\sigma_1^2, \cdots , \sigma_N^2]^\top$. The Maximum Likelihood Estimate of $\bbet$ for model \eqref{modelGLSP} is also the solution of the weighted least squares problem:

\begin{equation}
\label{eq:beta_wls}
\displaystyle{
\widehat{\bbet} :=  \arg \min_\bbet | \by - \bM \bbet     | ^2_{\boldsymbol{\Sigma}}
=  (\bM^\top  \boldsymbol{\Sigma} ^{-1} \bM)^{-1} \bM^\top \boldsymbol{\Sigma} ^{-1} \by}
\end{equation}

so that the fitted model is
\begin{equation}
\widehat{\by} := \bM \widehat{\bbet}   =  \bM  (\bM^\top  \boldsymbol{\Sigma} ^{-1} \bM)^{-1}  \bM^\top \boldsymbol{\Sigma} ^{-1} \by.
\end{equation}

and the error is
\begin{equation}
\widehat{{\bf{e}}}  := {\by}-\widehat{\by}    = \bM  (\bM^\top  \boldsymbol{\Sigma} ^{-1} \bM)^{-1} \bM^\top  \boldsymbol{\Sigma} ^{-1} \by.
\label{errors}
\end{equation}

To compare two models, say $\bM_1$ and $\bM_2$, a standard approach is to compare the corresponding residual sum of squares $\| \widehat{{\bf{e}}}_1\|^2$ and  $\| \widehat{{\bf{e}}}_2\|^2$,
where $ \widehat{{\bf{e}}}_1$   (resp. $ \widehat{{\bf{e}}}_2$) are computed by plugging $ \bM_1$ (resp. $\bM_2$) in place of $\bM$ in Eq. \eqref{errors}. A score $s$ can then be computed as:

\begin{equation}
s :=\displaystyle{\frac{\| \widehat{{\bf{e}}}_1\|^2-\| \widehat{{\bf{e}}}_2\|^2 }{ \| \widehat{{\bf{e}}}_1\|^2}}.
\end{equation}

When $\bM_1={\bf{1}}$ and $\bM_2=\bM_2(\nu)=[{\bf{1}}, \; \bcc(\nu), \; \bss(\nu)  ] $ as in Eq.~\ref{Mnu},
the resulting score $s=s(\nu)$ is the classical GLSP  $\cal{P}(\nu)$ \citep[Eq.(4) of ][]{zechmeister2009}.

With this description, it is straightforward to generalise this approach by including, for instance, a linear trend in the model, in which case
\begin{eqnarray}
    \bM_1&= &[{\bf{1}},\; {\bf{t}}]\\
    \bM_2&=&\bM_2(\nu)=[{\bf{1}}, \;{\bf{t}},\; \bcc(\nu), \; \bss(\nu)  ] 
\end{eqnarray}
or to account also for a quadratic trend, in which case
\begin{eqnarray}
    \bM_1&= &[{\bf{1}},\; {\bf{t}},\;{\bf{t}}^2]\; \\
    \bM_2&=&\bM_2(\nu)=[{\bf{1}}, \;{\bf{t}},\; {\bf{t}}^2,\; \bcc(\nu), \; \bss(\nu)  ] 
\end{eqnarray}
where ${\bf{t}}^2:=[t_1^2,\cdots,t_N^2]^\top$.
The columns can be normalised to improve numerical stability.

In Fig.~\ref{fig:sameGLSP}, we can see the results applied to the cases with and without a trend in the residuals. The periodograms shown in panel (a), obtained from the sample in panel (b), represent a typical example: in most cases, the frequency search results are comparable for the three models of the periodograms tested: a simple constant model, a model including a linear trend, and a model including linear and quadratic trends.

However, when the residuals present some trendy features, as in panel (d), we notice that the different periodograms provide different results, as shown in panel (c). Here, the largest peak that occurs at a low frequency for the classical GLSP is caused by the decreasing trend visible in panel (d). This is not the case for the two other periodograms, which are trend-insensitive. This makes it possible to detect the presence of potential oscillations at higher frequencies.

%****************************************************************
\section{Estimation performances at small amplitudes}
\label{app:performance}

\begin{figure}[htpb]
    \centering
    \includegraphics[width=0.95\linewidth]{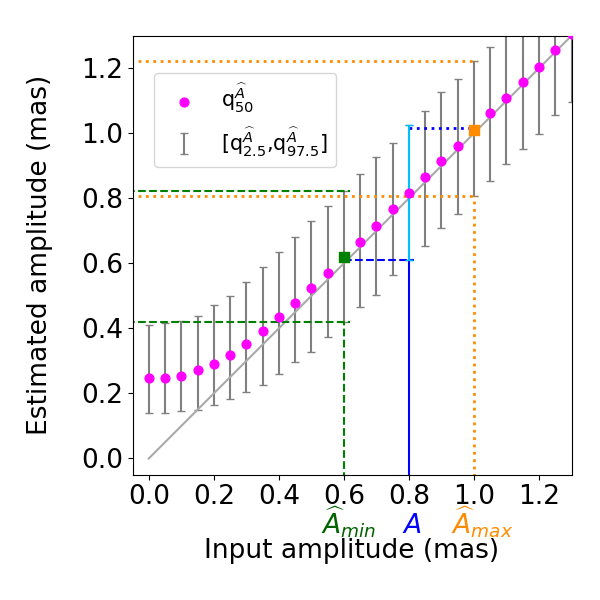}
    \caption{Wobble amplitude confidence interval estimation for a range of small amplitudes. The pink dots represent the quantiles $q^{\widehat{A}}_{50}$ estimated from the respective input amplitudes, while the error bars indicate the estimated confidence intervals $\mathcal{I}_{A}$= [$q^{\widehat{A}}_{2.5}$,$q^{\widehat{A}}_{97.5}$]. The coloured dotted and dashed lines show one example of a signal with true amplitude $A$= 0.8 mas, where $\mathcal{I}_{A}$= [0.61,1.02] mas, and where these limits are considered to be input values (${\widehat{A}_{min}}$ and ${\widehat{A}_{max}}$) from a signal with unknown amplitude (see text).}
    \label{fig:intervals}
\end{figure}

Figure \ref{fig:intervals} illustrates the  performances when estimating wobbles of amplitudes smaller than 1.3 mas. For each input amplitude value we obtain an estimated amplitude (pink dots) from the 50 quantile ($q^{\widehat{A}}_{50}$) and a corresponding confidence interval $\mathcal{I}_{A}$= [$q^{\widehat{A}}_{2.5}$,$q^{\widehat{A}}_{97.5}$], shown as the gray error bars. 

For a noisy sinusoidal signal, compatible with the \gfpr data set, where the nominal amplitude is $A$= 0.8 mas is shown by the blue solid line, the estimated amplitude $\widehat{A}$ can have any value in a confidence interval between $q^{\widehat{A}}_{2.5}\sim$0.61 and $q^{\widehat{A}}_{97.5}\sim$1.02 mas, shown as the light blue error bar.

We do not know the true amplitude of the wobble signatures in the \gaia astrometric data. So, in this example, we consider that the extreme cases of the confidence interval associated with $\widehat{A}$ are the input signal amplitude. If the signal is detected with an amplitude $\widehat{A}_{min}$=$q^{\widehat{A}}_{2.5}$ represented by the green square, the corresponding confidence interval is delimited as shown by the green horizontal dashed lines. 

The same observation can be made for the upper limit of the blue confidence interval where $\widehat{A}_{max}$= $q^{\widehat{A}}_{97.5}$, shown as the orange square, and the estimated confidence interval is delimited by the orange dotted lines. The true amplitude value $A$ is at the edges of the green and orange confidence intervals but still within both of them, showing that even when the true amplitude is unknown the method was capable of estimating confidence intervals that contain the true value. 

However, for amplitudes lower than 0.5 mas, the confidence intervals tend to be smaller and more asymmetrical, the 50\% quantiles tend to overestimate the amplitudes, and the confidence interval does not necessarily contain the true amplitude, explaining the larger false coverage rate obtained for amplitudes smaller than 1 mas, as observed in Fig.~\ref{fig:versions}.

\end{document}